\documentclass[prb,twocolumn,showpacs]{revtex4}
\usepackage{graphicx}

\catcode`\@=11
\def\simle{\mathrel{\mathpalette\@versim<}}   
\def\simge{\mathrel{\mathpalette\@versim>}}   
\def\@versim#1#2{\lower2.5pt\vbox{\baselineskip0pt \lineskip-.5pt
   \ialign{$\m@th#1\hfil##\hfil$\crcr#2\crcr\sim\crcr}}}
\newcommand{\mib}[1]{\mbox{\boldmath $#1$}} 

\begin{document}

\title{
Disorder Effect on Spin Excitation in Double Exchange Systems}

\author{Yukitoshi Motome}

\affiliation{
RIKEN (The Institute of Physical and Chemical Research), 
2-1 Hirosawa, Wako, Saitama 351-0198, Japan
}

\author{Nobuo Furukawa}

\affiliation{
Department of Physics, Aoyama Gakuin University, 
5-10-1 Fuchinobe, Sagamihara, Kanagawa 229-8558, Japan
}

\date{
\today
}

\begin{abstract}
Spin excitation spectrum is studied in the double exchange model
in the presence of disorder.
Spin wave approximation is applied in the lowest order of $1/S$ expansion.
The disorder causes anomalies in the spin excitation spectrum
such as broadening, branching, anticrossing with gap opening.
The origin of the anomalies is the Friedel oscillation,
in which the perfectly polarized electrons form the charge density wave 
to screen the disorder effect. 
Near the zone center $q = 0$,
the linewidth has a $q$ linear component
while the excitation energy scales to $q^2$,
which indicates that the magnon excitation is incoherent.
As $q$ increases, there appears a crossover 
from this incoherent behavior to the marginally coherent one 
in which both the linewidth and 
the excitation energy are proportional to $q^2$.
The results are compared with experimental results 
in colossal magnetoresistance manganese oxides. 
Quantitative comparison of the linewidth suggests that 
spatially-correlated or mesoscopic-scale disorder is more relevant
in real compounds than local or atomic-scale disorder. 
Comparison with other theoretical proposals is also discussed.
Experimental tests are proposed for the relevance of disorder.
\end{abstract}


\maketitle

\section{Introduction}
\label{Sec:Introduction}

Since rediscovery of the colossal magnetoresistance (CMR) phenomena,
perovskite manganese oxides have attracted much attention.
\cite{Ramirez1997,Coey1999,Salamon2001,Imada1998,Tokura1999,Tokura2000,Dagotto2001}
This family of compounds shows a variety of physical properties
in magnetic (ferromagnetism or antiferromagnetism),
charge (metal, insulator, or charge ordering), and
orbital/lattice (orbital ordering and/or Jahn-Teller distortion) 
degrees of freedom.
Phase transitions between different ordered phases can be controlled
by chemical substitution of the {\it A} site
in the general chemical formula {\it A}MnO$_3$.
The magnitude of the CMR effect also changes drastically
for different {\it A}-site ions, hence,
one of the most important issues in this field is
to understand the role of the {\it A}-site substitution.

The ferromagnetic metallic state,
which is widely stable at low temperatures in these compounds,
\cite{Wollan1955,Searle1970}
is basically understood by the Zener's double exchange (DE) mechanism.
\cite{Zener1951,Anderson1955,deGennes1960}
The kinetics of itinerant electrons in the Mn $e_g$ bands
strongly correlates with localized spins in the $t_{2g}$ levels
through the large Hund's-rule coupling;
ferromagnetism of the localized spins leads to
high conductivity of the electrons, and vice versa.
Thus, the issue is what is an additional element to the DE interaction
which is necessary to explain deviations from this canonical DE behavior and
instabilities towards a variety of different phases 
in the {\it A}-site substituted materials. 

Among various manganese oxides, it has been recognized that
La$_{1-x}$Sr$_x$MnO$_3$ (LSMO) at $x \sim 0.3$ is a canonical DE system.
\cite{Furukawa1999a}
Namely, the simple DE model can explain magnetic, transport
and optical properties of this material quantitatively.
For instance, the experimental value of 
the Curie temperature $T_{\rm C}$
is well reproduced by recent theoretical calculations in the DE model.
\cite{Motome2000,MotomePREPRINT}
This material shows relatively high $T_{\rm C}$
compared to other manganese oxides, 
which also suggests
the stability of ferromagnetic metal due to
the primary role of the DE mechanism and
the irrelevance of other additional factors.
This canonical DE system gives a good starting point
to examine effects of the {\it A}-site substitution.

There are two different ways of the systematic {\it A}-site substitution.
One is the substitution with different ionic valences,
such as the control of $x$ in La$_{1-x}$Sr$_x$MnO$_3$.
The mixture of trivalent (La$^{3+}$) and divalent (Sr$^{2+}$) ions 
effectively changes the nominal valence of manganese ion 
as Mn$^{(3+x)+}$,
which controls the electron density in the Mn-O network.
The other substitution is by different ionic radii at a fixed valence,
such as {\it A}$_{1-x}${\it A}'$_{x}$MnO$_{3}$ 
for different combinations of, for instance, 
{\it A}=La,Pr,Nd,Y and {\it A}'=Ba,Sr,Ca at a fixed $x$.
These two different substitutions cause
systematic and drastic deviations from the canonical DE behavior
in LSMO compounds at $x \sim 0.3$.

We discuss mainly the latter {\it A}-site substitution
with the ionic-radius control in the following.
The change of the ionic radii is a sort of the chemical pressure,
which modifies length and angle of Mn-O-Mn bonds.
This leads to the change of effective transfer integrals between Mn ions,
namely, the bandwidth of the electrons.
Thus, this substitution has been often called the bandwidth control.
This terminology `bandwidth control', however, might be misleading:
There are many experimental facts
which cannot be explained only by the change of the bandwidth.
One is a rapid decrease of $T_{\rm C}$ compared to the bandwidth change.
\cite{Hwang1995,Radaelli1997}
$T_{\rm C}$ decreases as the averaged radius of the {\it A} ions decreases;
however, the change of $T_{\rm C}$ is much larger than 
that of the bandwidth which is estimated from the structural change.
For instance, from La$_{0.7}$Sr$_{0.3}$MnO$_3$
to La$_{0.7}$Ca$_{0.3}$MnO$_3$ (LCMO),
$T_{\rm C}$ decreases by about $30$\% 
while the estimated bandwidth decreases by less than $2$\%.
\cite{Radaelli1997}
Since, in the DE theory, $T_{\rm C}$ is proportional to the bandwidth,
the large change of $T_{\rm C}$ cannot be explained by the bandwidth change alone.
There should be a hidden parameter 
which strongly suppresses the kinetics of the electrons.

In this work, we focus on the quenched disorder for a candidate of
the hidden parameter in this {\it A}-site substitution.
In general, oxides are known to be far from perfect crystals.
Moreover, in these manganese oxides, the disorder is inevitably introduced
since they are solid solutions of different {\it A}-site ions.

The importance of the disorder has been pointed out
in several experimental results:
(i) $T_{\rm C}$ changes not only as the average of the ionic radius
but also as its standard deviation.
\cite{Rodriguez-Martinez1996}
Even if the average is the same, 
$T_{\rm C}$ becomes lower for larger standard deviation.
(ii) The residual resistivity, namely, the value of the resistivity 
extrapolated to zero temperature, becomes larger for lower-$T_{\rm C}$ compounds.
\cite{Coey1995,Saitoh1999}
(iii) {\it A}-site ordered materials, in which different {\it A} ions
form a periodic ordered structure, 
exhibit higher $T_{\rm C}$ compared to 
compounds with random distribution of {\it A} ions
in the same chemical formula.
\cite{Millange1998,Nakajima2002,Akahoshi2003}
All these experimental results suggest that 
the disorder in the random mixture of different size ions
scatters the itinerant electrons and
suppresses the kinetics of them.

In this paper, we discuss the disorder effect on spin dynamics
in the {\it A}-site substituted manganites.
The spin dynamics is 
another important indication of a hidden parameter.
Spin excitation spectrum shows qualitative changes
for the {\it A}-site substitution.
In compounds with relatively high $T_{\rm C}$ such as LSMO,
the spin excitation shows a cosine type dispersion, 
which is similar to that of Heisenberg spin systems.
\cite{Perring1996,Moudden1998}
This behavior is well described by the DE mechanism alone.
\cite{Furukawa1996}
On the contrary, in compounds with low $T_{\rm C}$ such as LCMO,
the spectrum shows significant deviations 
from this form, e.g., some anomalies
such as broadening, softening, and gap opening.
\cite{Hwang1998,Vasiliu-Doloc1998,Dai2000,Biotteau2001}
To explain these anomalies
is one of the crucial test for elucidating the hidden parameter.

We will demonstrate here that the disorder gives 
a comprehensive understanding of
systematic changes of the spin excitation spectrum as well as
other experimental results described above.
Several other mechanisms have been proposed,
for instance, 
antiferromagnetic interactions between localized spins,
\cite{Solovyev1999}
orbital degrees of freedom in the $e_g$ bands,
\cite{Khaliullin2000}
the electron-lattice coupling, 
\cite{Furukawa1999b}
and the electron-electron correlation.
\cite{Kaplan1997,Golosov2000,Shannon2002}
Through the detailed analysis of the spectrum,
we will show that the disorder appears to be promising
among these scenarios.
A part of the results has been published in the previous publications.
\cite{Motome2002a,FurukawaPREPRINT,Motome2003}
More systematic and extensive analyses are presented in this paper.

This paper is organized as follows.
In Sec.~\ref{Sec:Formulation}, we describe
methods to study spin dynamics in the DE model.
To be self-contained, we derive and summarize
some analytical expressions which have been developed
in the previous publications.
\cite{Motome2002a,FurukawaPREPRINT,Motome2003}
Numerical results are presented in Sec.~\ref{Sec:Results}
for wide regions of parameters. 
We find remarkable anomalies in the spin excitation spectra and
discuss their origin in detail.
Comparison with the analytical results is also examined.
The results are compared with
experimental results as well as with other theoretical ones
in Sec.~\ref{Sec:Discussions}.
We propose possible experiments to test the relevance of the disorder.
Section \ref{Sec:Summary} is devoted to summary and concluding remarks.

\section{Formulation}
\label{Sec:Formulation}

\subsection{Model}
\label{Sec:Model}

In this work, we study effects of the disorder
on the spin dynamics of the DE model.
\cite{Zener1951,Anderson1955}
Our Hamiltonian is given in the form
\begin{eqnarray}
{\cal H} = &-& \sum_{i<j} \sum_{\sigma = \pm} t_{ij} ( 
c_{i\sigma}^\dagger c_{j\sigma} + {\rm h.c.} )
- J_{\rm H} \sum_i \mib{\sigma}_i \cdot
\mib{S}_i 
\nonumber \\
&+& \sum_{i\sigma} \varepsilon_i c_{i\sigma}^\dagger c_{i\sigma}.
\label{eq:H_DE}
\end {eqnarray}
Here, the first term describes the electron hopping 
between $i$ and $j$th site,
the second term represents the Hund's-rule coupling
between the electron spin {\boldmath $\sigma$}
(vector of Pauli matrix) and the localized spin {\boldmath $S$},
and the last term contains the on-site potential $\varepsilon_i$. 

In model (\ref{eq:H_DE}), we take account of the quenched disorder in two ways.
One is the off-diagonal disorder in the transfer integral $t_{ij}$ and
the other is the diagonal disorder in the potential energy $\varepsilon_i$.
Hereafter, we call the former the bond disorder and
the latter the on-site disorder, respectively.
The following discussions in this Sec.~\ref{Sec:Formulation}
do not depend on the detailed functional form of 
the distribution of the random variables $t_{ij}$ and $\varepsilon_i$. 
The distribution will be specified in Sec.~\ref{Sec:Randomness} 
for numerical calculations in Sec.~\ref{Sec:Results}.

\subsection{$1/S$ expansion}
\label{Sec:1/S}

We apply the spin wave approximation
in the lowest order of $1/S$ expansion 
to model (\ref{eq:H_DE})
($S$ is the magnitude of the localized spin {\boldmath $S$}).
We consider a ferromagnetic ground state
with perfect spin polarization $\mib{S}_i = (0,0,S)$,
\cite{t_ij}
and calculate the one-magnon excitation spectrum.
In the absence of disorder,
the formulation is given in Ref.~\onlinecite{Furukawa1996}.
Here, we extend the method to the case with disorder.

\begin{figure}
\includegraphics[width=7cm]{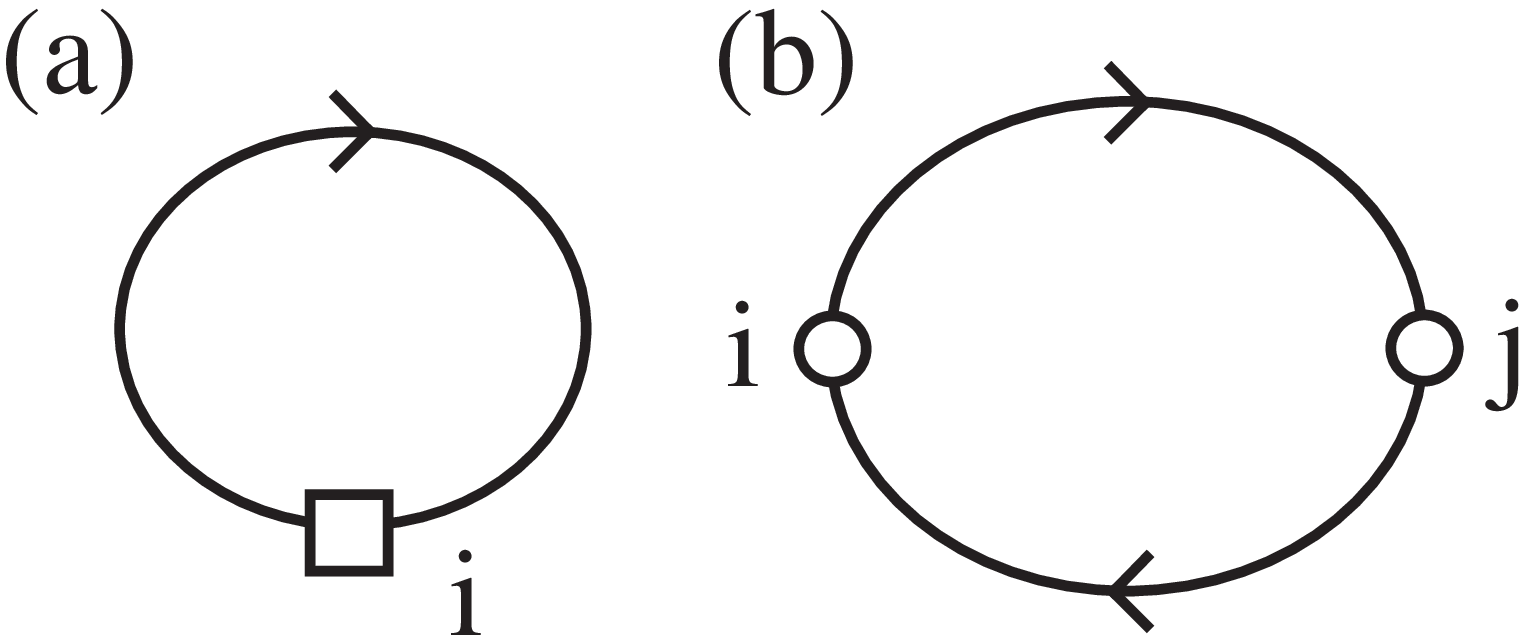}
\caption{
Spin wave self-energy in the lowest order of the $1/S$ expansion.
See Ref.~\onlinecite{Furukawa1996} for details.
}
\label{fig:Pi_ij}
\end{figure}

In the presence of disorder,
it is convenient to work with the real-space representation
instead of the momentum-space one.
First, we explicitly diagonalize the Hamiltonian matrix
of model~(\ref{eq:H_DE}) for a given configuration of the disorder
$\{t_{ij},\varepsilon_i\}$.
We have
\begin{equation}
\sum_{j} {\cal H}_{ij} (\{t_{ij},\varepsilon_i\}) \varphi_{n\sigma}(j)
= (E_n - \sigma J_{\rm H}) \varphi_{n\sigma}(i),
\label{eq:eigenH_DE}
\end{equation}
where $\varphi_{n\sigma}$ is the $n$th eigenstate 
with the eigenenergy $E_n$.
Electron Green's function is given by
\begin{equation}
G_{ij,\sigma}(\omega) = \sum_n
\frac{\varphi_{n\sigma}(i) \varphi_{n\sigma}^*(j)}
{\omega - (E_n - \sigma J_{\rm H} - \mu) + {\rm i} \eta \ 
{\rm sgn}\omega},
\label{eq:G_ij}
\end{equation}
where $\mu$ is the chemical potential and 
$\eta$ is an infinitesimal for convergence. 
In the lowest order of the $1/S$ expansion, 
following Ref.~\onlinecite{Furukawa1996},
the magnon self-energy is obtained 
from the electron spin polarization function
shown in Fig.~\ref{fig:Pi_ij}.
The contribution from Fig.~\ref{fig:Pi_ij} (a) is given by
\begin{equation}
\Pi_{ii}^z(\omega) = \frac{J_{\rm H}}{S}
\sum_n f_{n+} |\varphi_{n+}(i)|^2,
\label{eq:Pi_ij_a}
\end{equation}
where $f_{n+}$ is the fermi distribution function for up-spin states.
The other contribution from Fig.~\ref{fig:Pi_ij} (b) is given by
\begin{equation}
\Pi_{ij}^{xy}(\omega) = \frac{2J_{\rm H}^2}{S}
\sum_{mn} f_{n+} \frac{\varphi_{n+}(i) \varphi_{n+}^*(j) 
\varphi_{m+}(j) \varphi_{m+}^*(i)}
{\omega + E_n - E_m - 2J_{\rm H}}.
\label{eq:Pi_ij_b}
\end{equation}
In the limit of $J_{\rm H} \gg t$, 
which is realistic in CMR manganites,
the summation of Eqs.~(\ref{eq:Pi_ij_a}) and (\ref{eq:Pi_ij_b})
ends up with
\begin{eqnarray}
\Pi_{ij}(\omega) &=& \frac{1}{2S}
\sum_{mn} f_n \varphi_n(i) \varphi_n^*(j) \varphi_m(j) \varphi_m^*(i)
\nonumber \\
&& \quad \quad \quad \quad \ \ \times
(E_m - E_n - \omega).
\label{eq:Pi_ij_sum}
\end{eqnarray}
Here, we assume the orthonormal relation 
\begin{equation}
\sum_m \varphi_m(j) \varphi_m(i) = \delta_{ij},
\label{eq:orthonormal}
\end{equation} 
and drop the spin index for simplicity.

Magnon Green's function is calculated by this self-energy in the form 
\begin{equation}
D_{ij}(\omega) = \left[ \ \omega - \Pi_{ij}(\omega) 
+ {\rm i}\eta \ \right]^{-1}.
\label{eq:D_ij}
\end{equation}
Spin wave excitations are obtained from the poles of 
the Green's function (\ref{eq:D_ij}).
Since the self-energy $\Pi_{ij}$ in Eq.~(\ref{eq:Pi_ij_sum})
is proportional to $1/S$,
the positions of the poles are at $\omega \sim O(1/S)$.
Hence, within the lowest order of $1/S$ expansion,
spin wave excitation is determined
by the static part of the self-energy $\Pi_{ij}(\omega=0)$
as discussed in Ref.~\onlinecite{Furukawa1996}.
The matrix inversion in Eq.~(\ref{eq:D_ij}) is easily obtained
by using the eigenvectors $\psi_l$ and the eigenvalues $\omega_l$
which satisfy
\begin{equation}
\sum_j \Pi_{ij}(\omega=0) \cdot \psi_l(j) = \omega_l \psi_l(i).
\label{eq:eigenPi}
\end{equation}

Finally, the spectral function for a given configuration of
$\{t_{ij},\varepsilon_i\}$ is calculated as
\begin{eqnarray}
A(\mib{q},\omega) &=&
- \frac1N \sum_{ij} \frac{1}{\pi} {\rm Im} D_{ij}(\omega)
{\rm e}^{{\rm i} \mib{q} \cdot (\mib{r}_i - \mib{r}_j)}
\nonumber \\
&=& \frac1N \sum_l \Big| \sum_j \psi_l(j)
{\rm e}^{{\rm i} \mib{q} \cdot \mib{r}_j} 
\Big|^2 \delta(\omega - \omega_l),
\label{eq:Aqw}
\end{eqnarray}
where $N$ is the system size.
The spin excitation spectrum is obtained by averaging $A(\mib{q},\omega)$
for random configurations of $\{t_{ij},\varepsilon_i\}$. 
In the same manner, all the physical quantities are 
to be averaged for random configurations 
although we do not explicitly remark in the following discussions.

To summarize, the spin excitation spectrum in the DE model 
is calculated up to the lowest order of the $1/S$ expansion
as follows.
(i) Diagonalize the Hamiltonian (\ref{eq:H_DE})
for a configuration of the disorder $\{t_{ij},\varepsilon_i\}$,
and obtain the electron wavefunction $\varphi_n$ and 
eigenenergy $E_n$.
(ii) Calculate the magnon self-energy (\ref{eq:Pi_ij_sum})
by using $\varphi_n$ and $E_n$.
(iii) Diagonalize the self-energy as in Eq.~(\ref{eq:eigenPi}),
and obtain the eigenvalues $\omega_l$ and eigenvectors $\psi_l$.
(iv) Calculate $A(\mib{q},\omega)$ according to Eq.~(\ref{eq:Aqw}).
(v) Repeat (i)-(iv) for different random configurations
and take the random average of $A(\mib{q},\omega)$.

\subsection{Analytical formula}
\label{Sec:Analytical}

In this section, 
we derive some analytical expressions for the spin wave excitation.
The following framework is general and 
does not depend on the details of the disorder and 
the spatial dimension of the system.
The derived formulas are useful to discuss 
the numerical data in the following sections.

\subsubsection{Magnon self-energy}
\label{Sec:self-energy}

First, we examine the static part of the magnon self-energy.
We can rewrite $\Pi_{ij}(\omega=0)$ in the form
\begin{equation}
\Pi_{ij}(0) = \frac{1}{2S}
\Big( {\cal H}_{ij} B_{ji} - \delta_{ij} \sum_k
{\cal H}_{ik} B_{ki} \Big), 
\label{eq:Pi_ij_by_B_ij}
\end{equation}
where $B_{ij} = \sum_n f_n \varphi_n(j) \varphi_n^*(i)$.
Here, we use the following relations
\begin{eqnarray}
\sum_m E_m \varphi_m(i) \varphi_m^*(j) &=&
\sum_{mk} {\cal H}_{ik} \varphi_m(k) \varphi_m^*(j) 
\nonumber \\
&=& {\cal H}_{ij},
\\
\sum_n f_n E_n \varphi_n(j) \varphi_n^*(i) &=&
\sum_{nk} f_n {\cal H}_{jk} \varphi_n(k) \varphi_n^*(i) 
\nonumber \\
&=& \sum_k {\cal H}_{jk} B_{ki},
\end{eqnarray}
with the eigenequation (\ref{eq:eigenH_DE}) and 
the orthonormality (\ref{eq:orthonormal}).

From Eq.~(\ref{eq:Pi_ij_by_B_ij}),
we can easily show that the self-energy satisfies the sum rule 
in the form
\begin{equation}
\sum_j \Pi_{ij}(0) = \frac{1}{2S}
\Big( \sum_j {\cal H}_{ij} B_{ji} - \sum_k {\cal H}_{ik} B_{ki} \Big)
= 0.
\label{eq:sum rule}
\end{equation}
Thus, the eigenvalue in Eq.~(\ref{eq:eigenPi}) always becomes
zero at $\mib{q} = 0$ even in the presence of disorder.
This indicates that there is a gapless excitation at $\mib{q} = 0$.
This is consistent with the Goldstone theorem.

We note also that the matrix element of $\Pi_{ij}(0)$ 
is given by the transfer energy of the electrons.
By using the relation
\begin{equation}
{\cal H}_{ij} B_{ji} = \sum_n f_n \varphi_n^*(i) {\cal H}_{ij}
\varphi_n(j) = \langle {\cal H}_{ij} \rangle,
\label{eq:H_ij_B_ij}
\end{equation}
the matrix elements are written as
\begin{eqnarray}
&& \Pi_{i \neq j}(0) = - t_{ij}
\langle c_i^\dagger c_j \rangle / 2S
\equiv -2S J_{ij},
\label{eq:Pi_ij_by_J_ij}
\\
&& \Pi_{ii}(0) = 2S \sum_j J_{ij},
\label{eq:Pi_ii_by_J_ij}
\end{eqnarray}
where we define
\begin{equation}
J_{ij} = t_{ij} \langle c_i^\dagger c_j \rangle / 4S^2.
\label{eq:J_ij}
\end{equation}
The bracket represents the expectation value in the ground state
for a given configuration of the disorder.
We will show in Sec.~\ref{Sec:Heisenberg} that 
$J_{ij}$ is the exchange coupling
of the corresponding Heisenberg spin model.
Equations~(\ref{eq:Pi_ij_by_J_ij}) and (\ref{eq:Pi_ii_by_J_ij})
indicate that
the following summations give the kinetic energy of electrons;
\begin{equation}
\sum_{i \neq j} \Pi_{ij}(0) = -\sum_i \Pi_{ii}(0)
= \langle {\cal H}_{\rm kin} \rangle / 2S,
\label{eq:sum rule to Ekin}
\end{equation}
where ${\cal H}_{\rm kin}$ is the first term in the Hamiltonian (\ref{eq:H_DE}).

From Eqs.~(\ref{eq:Pi_ij_by_J_ij}) and (\ref{eq:Pi_ii_by_J_ij}),
we note that the magnon self-energy at $\omega=0$ is real in general
since $\langle c_i^\dagger c_j \rangle$ is real
for the fully-polarized ferromagnetic ground state
without degeneracy.
Hence, as seen in Eqs.~(\ref{eq:eigenPi}) and (\ref{eq:Aqw}),
up to $O(1/S)$, $A(\mib{q},\omega)$ 
for a given configuration of disorder
describes a well-defined quasi-particle excitation
with infinite lifetime (zero linewidth).
Lifetime due to the decay of the quasi particle
is obtained from the imaginary part of
the poles of Eq.~(\ref{eq:D_ij}), however,
it is a higher order term in the $1/S$ expansion.
(This higher order correction will be discussed 
in Sec.~\ref{Sec:Broadening}.)
On the other hand, in the presence of disorder,
we have to take the random average 
for different realizations of disorder configurations.
Since different configurations give different $A(\mib{q},\omega)$,
we obtain a distribution of the excitation spectra 
after the random average.
The distribution gives rise to a finite linewidth,
which can be estimated, for instance, 
by the standard deviation of the excitation energy.
This linewidth is due to the so-called inhomogeneous broadening.
In the presence of the disorder,
this broadening is the lowest order effect in the $1/S$ expansion.
We will analyze this linewidth in the following sections.

\subsubsection{Spectral function analysis}
\label{Sec:Analysis}

If the spin excitation spectrum is single-peaked and 
does not show any splitting,
we can apply the following spectral function analysis.
We will show in Sec.~\ref{Sec:Results} that 
this is the case near the zone center $\mib{q} = 0$.
Let us consider the $m$th moment of the spectral function as
\begin{equation}
\Omega_{\mib{q}}^{(m)} =
\int_0^\infty \omega^m A(\mib{q},\omega) \ d\omega.
\label{eq:Omega^m}
\end{equation}
Using Eqs.~(\ref{eq:eigenPi}) and (\ref{eq:Aqw}), we obtain
\begin{equation}
\Omega_{\mib{q}}^{(m)} = \frac1N
\sum_{ij} \sum_{k_1 k_2 ...k_{m-1}}
\Pi_{ik_1} \Pi_{k_1 k_2} \cdot \cdot \cdot
\Pi_{k_{m-1} j} 
{\rm e}^{{\rm i} \mib{q} \cdot (\mib{r}_i - \mib{r}_j)}.
\label{eq:Omega^m_by_Pi_ij}
\end{equation}
Thus the moment $\Omega_{\mib{q}}^{(m)}$
is a Fourier transform of the magnon self-energy $(\Pi_{ij}(0))^m$.

The first moment in Eq.~(\ref{eq:Omega^m}) gives
the averaged excitation energy as
\begin{equation}
\omega_{\rm sw} (\mib{q}) = \Omega_{\mib{q}}^{(1)}.
\label{eq:omega_sw}
\end{equation}
As mentioned in the last part of Sec.~\ref{Sec:self-energy},
the linewidth is estimated by the standard deviation of the excitation energy 
which is given by the second moment in the form
\begin{equation}
\gamma (\mib{q}) =
\big[ \ \Omega_{\mib{q}}^{(2)} - (\Omega_{\mib{q}}^{(1)})^2 \ \big]^{1/2}.
\label{eq:gamma}
\end{equation}

\subsubsection{Excitation energy}
\label{Sec:Excitation energy}

Here, we analyze the excitation energy of Eq.~(\ref{eq:omega_sw}).
Substituting Eqs.~(\ref{eq:Omega^m}) and (\ref{eq:Omega^m_by_Pi_ij})
into Eq.~(\ref{eq:omega_sw}), we obtain
\begin{equation}
\omega_{\rm sw} (\mib{q}) =
\frac1N \sum_{ij} \Pi_{ij} {\rm e}^{{\rm i} \mib{q} \cdot
(\mib{r}_i - \mib{r}_j)}.
\end{equation}
When model~(\ref{eq:H_DE}) has the electron hopping 
only between the nearest neighbor sites, 
the magnon self-energy $\Pi_{ij}$ has nonzero matrix elements
only for $i=j$ and $i=j+\mib{\eta}$
as shown by Eqs.~(\ref{eq:Pi_ij_by_J_ij}) and (\ref{eq:Pi_ii_by_J_ij}).
Here $\mib{\eta}$ is a displacement vector
to the nearest neighbor site.
Then the excitation energy is given as
\begin{eqnarray}
\omega_{\rm sw} (\mib{q}) &=& \frac1N \sum_i
\Big( \Pi_{ii} + \sum_{\mib{\eta}}
\Pi_{i,i+\mib{\eta}} \ 
{\rm e}^{{\rm i} \mib{q} \cdot \mib{\eta}} \Big)
\nonumber \\
&=& \sum_{\mib{\eta}} \ \bar{\Pi} (\mib{\eta}) \ 
({\rm e}^{{\rm i} \mib{q} \cdot \mib{\eta}} - 1),
\label{eq:omega_sw_cos}
\end{eqnarray}
where we define the site-averaged quantity
$\bar{\Pi} (\mib{\eta}) = 
\sum_i \Pi_{i,i+\mib{\eta}} / N$ and
use the sum rule Eq.~(\ref{eq:sum rule}).
Even in the presence of the disorder,
the symmetry for the direction of $\mib{\eta}$
is expected to be recovered after the random average,
therefore we assume
\begin{equation}
\bar{\Pi} (\mib{\eta}) = \frac1z \sum_{\mib{\eta}}
\bar{\Pi} (\mib{\eta}) \equiv - \Lambda,
\label{eq:Lambda}
\end{equation}
where $z$ is the number of coordinates.
Finally, we obtain the following form
\begin{equation}
\omega_{\rm sw} (\mib{q}) = 
\Lambda \sum_{\mib{\eta}}
(1 - {\rm e}^{{\rm i} \mib{q} \cdot \mib{\eta}}).
\label{eq:omega_sw_final}
\end{equation}
This indicates that on hypercubic lattices
the excitation spectrum shows 
the cosine dispersion even in the presence of disorder.
In the limit of $q = | \mib{q} | \rightarrow 0$,
we have
\begin{equation}
\omega_{\rm sw} (\mib{q}) \simeq \Lambda q^2.
\label{eq:omega_sw_q->0}
\end{equation}
Thus, the quantity $\Lambda$ gives the spin wave stiffness.
In our analysis, the stiffness is proportional to 
the kinetic energy of electrons as
\begin{equation}
\Lambda = -\frac1z \sum_{\mib{\eta}}
\bar{\Pi} (\mib{\eta})
= -\frac{1}{zN} \sum_{i \mib{\eta}}
\Pi_{i,i+\mib{\eta}}
= -\frac{\langle {\cal H}_{\rm kin} \rangle}{2SzN},
\label{eq:Lambda to Ekin}
\end{equation}
where we use Eq.~(\ref{eq:sum rule to Ekin}).

\subsubsection{Linewidth}
\label{Sec:Linewidth}

Next, we analyze the linewidth of Eq.~(\ref{eq:gamma}).
Similarly to the derivation of Eq.~(\ref{eq:omega_sw_cos}),
the second moment of the spectral function is written in the form
\begin{equation}
\Omega_{\mib{q}}^{(2)} =
\sum_{\mib{\eta}_1 \mib{\eta}_2} \
\bar{\Pi}_2 (\mib{\eta}_1,\mib{\eta}_2)
(1 - {\rm e}^{{\rm i} \mib{q} \cdot \mib{\eta}_1})
(1 - {\rm e}^{{\rm i} \mib{q} \cdot \mib{\eta}_2}),
\label{eq:Omega_q^2}
\end{equation}
where
\begin{equation}
\bar{\Pi}_2 (\mib{\eta}_1,\mib{\eta}_2)
= \frac1N \sum_i \Pi_{i+\mib{\eta}_1,i}
\Pi_{i,i+\mib{\eta}_2},
\label{eq:barPi_2}
\end{equation}
and $\mib{\eta}_1$ and $\mib{\eta}_2$ 
are the displacement vectors to the nearest neighbor site.
In the absence of disorder, we have
$\Pi_{i+\mib{\eta},i} = \Pi_{i,i+\mib{\eta}} = - \Lambda$.
Hence, the linewidth $\gamma$ becomes zero
since $\Omega_{\mib{q}}^{(2)} = ( \Omega_{\mib{q}}^{(1)} )^2$.
In the presence of disorder, the linewidth becomes finite.
For simplicity, we consider here
the linewidth in a special direction
such as $\mib{q} = (q,0,0, \cdot \cdot \cdot)$
on a hypercubic lattice.
Since $\bar{\Pi}_2$ in Eq.~(\ref{eq:barPi_2}) generally depends on
the relative direction between $\mib{\eta}_1$ and $\mib{\eta}_2$,
we represent two different cases as
\begin{equation}
\bar{\Pi}_2 (\mib{\eta}_1,\mib{\eta}_2) = 
\Lambda_2 \pm \delta \Lambda_2 \quad 
\mbox{for \ $\mib{\eta}_1 = \pm \mib{\eta}_2$}.
\end{equation}
Then, Eq.~(\ref{eq:Omega_q^2}) is rewritten as
\begin{equation}
\Omega_{\mib{q}}^{(2)} =
4 \big( \ \Lambda_2 (1 - \cos q )^2
+ \delta \Lambda_2 \sin^2 q \ \big).
\end{equation}
In the limit of $q \rightarrow 0$, we have
\begin{equation}
\gamma^2 (\mib{q}) \simeq c_1 q^2 + c_2 q^4,
\label{eq:gamma_q->0}
\end{equation}
where
\begin{eqnarray}
c_1 &=& 4 \ \delta \Lambda_2, 
\label{eq:c_1}
\\
c_2 &=& \Lambda_2 - \frac43 \ \delta \Lambda_2 - \Lambda^2.
\label{eq:c_2}
\end{eqnarray}
Therefore, in the presence of disorder,
the linewidth shows a $q$ linear behavior in the small $q$ regime.

\subsubsection{Incoherence of spin wave excitation}
\label{Sec:Incoherence}

From the above analysis,
we obtain $\omega_{\rm sw} \propto q^2$ and
$\gamma \propto q$ in the limit of $q \rightarrow 0$
as in Eqs.~(\ref{eq:omega_sw_q->0}) and (\ref{eq:gamma_q->0}).
Thus, the linewidth $\gamma$ becomes larger than
the excitation energy $\omega_{\rm sw}$ at $\mib{q} \sim 0$, 
which indicates that the spin wave excitation 
becomes incoherent or localized.

This incoherent behavior comes from local fluctuations.
The $q$ linear coefficient in $\gamma$ in Eq.~(\ref{eq:c_1})
is given by the difference of $\bar{\Pi}_2$
for different directions of $\mib{\eta}$,
which is proportional to
\begin{eqnarray}
c_1 &\propto&
\bar{\Pi}_2 (\mib{\eta},\mib{\eta}) - \bar{\Pi}_2  (\mib{\eta},-\mib{\eta})
\nonumber \\
&\propto&
\sum_i ( \Pi_{i+\mib{\eta},i} - \Pi_{i,i-\mib{\eta}} )^2
\nonumber \\
&\propto&
\sum_i ( J_{i+\mib{\eta},i} - J_{i,i-\mib{\eta}} )^2,
\label{eq:c_1_by_J_ij}
\end{eqnarray}
where we use Eq.~(\ref{eq:Pi_ij_by_J_ij}).
From the definition of $J_{ij}$ in Eq.~(\ref{eq:J_ij}),
Eq.~(\ref{eq:c_1_by_J_ij}) shows that 
the $q$ linear coefficient $c_1$ is proportional to 
the local fluctuations of the transfer energy of electrons.

As $q$ becomes large,
we have a crossover from this incoherent regime
where $\gamma \propto \omega_{\rm sw}^{1/2} \propto q$
to the marginally coherent regime 
where $\gamma \propto \omega_{\rm sw} \propto q^2$.
This crossover is determined by 
the coefficients $c_1$ and $c_2$ in Eqs.~(\ref{eq:c_1}) and (\ref{eq:c_2}).
These coefficients depend on the details of parameters in the system,
especially on the type of the disorder, 
as will be discussed in Sec~\ref{Sec:omega and gamma}.

\subsection{Correspondence to Heisenberg model}
\label{Sec:Heisenberg}

Here, we discuss the relation between
the spin excitation spectra in the DE model (\ref{eq:H_DE})
and those in the Heisenberg spin model
whose Hamiltonian is given by
${\cal H}_{\rm Heis} = -2 \sum_{i<j} J_{ij} \mib{S}_i \cdot \mib{S}_j$.
As explicitly shown in Appendix~\ref{AppendixA},
the static part of the magnon self-energy 
(\ref{eq:Pi_ij_by_B_ij}) corresponds to
the effective spin-wave Hamiltonian for the Heisenberg model
within the leading order of the $1/S$ expansion.
There, Eq.~(\ref{eq:J_ij}) gives
the relation between
the kinetics of itinerant electrons in the DE model (\ref{eq:H_DE}) 
and the exchange coupling $J_{ij}$ in the Heisenberg model.
This relation provides us
physical intuitions to understand our results.

In the absence of disorder, 
when we consider the transfer $t_{ij} = t$ 
only for the nearest neighbor sites,
$J_{ij}$ in Eq.~(\ref{eq:J_ij}) becomes a constant $J$
which is nonzero only for $i = j + \mib{\eta}$.
Thus, as noted in Ref.~\onlinecite{Furukawa1996},
the spin excitation spectrum in the pure DE model with infinite $J_{\rm H}$
is equivalent to that in the Heisenberg model 
with the nearest neighbor exchange $J_{ij} = J$, and
is given by the simple cosine dispersion.

Disorder introduces some randomness in the expectation value
of $\langle c_i^\dagger c_j \rangle$ as well as
the transfer integral $t_{ij}$.
This leads to the nonuniform exchange coupling $J_{ij}$
in the corresponding Heisenberg model.
It is instructive to compare the spectrum for the DE model and
that for the Heisenberg model with random exchange couplings $J_{ij}$
which no longer satisfy the relation (\ref{eq:J_ij}).
(See also Sec.~\ref{Sec:Random Heisenberg}.)
For the random Heisenberg model,
the spin wave stiffness $\Lambda$ 
is given by the random average of the exchange constant $J_{ij}$.
Moreover, the $q$ linear term
in the linewidth is determined by
local fluctuations of the random exchange $J_{ij}$.
Thus, these aspects are commonly observed
both in the random DE model and in the random Heisenberg model
even if the relation (\ref{eq:J_ij}) does not hold
between them.
In other words, in the long wavelength limit of $q \rightarrow 0$,
the magnetic excitation has a similar structure
between these two models, 
although one is the itinerant electron system and 
the other is the localized spin system.
However, for moderate or large $q$,
the excitation spectrum of the DE model in the presence of disorder 
shows remarkable differences from that of the random Heisenberg model
as shown in Sec.~\ref{Sec:Results}.
There, the fermionic properties of itinerant electrons 
play a key role through the expectation value
$\langle c_i^\dagger c_j \rangle$.

\section{Results}
\label{Sec:Results}

\subsection{Overview}
\label{Sec:Overview}

In this section, we show numerical results which are obtained
based on the method described in Sec.~\ref{Sec:1/S}.
After the specification of the functional form of the disorder distribution 
in Sec.~\ref{Sec:Randomness}, 
we show that the spin excitation spectrum
exhibits some anomalies due to the disorder
in Sec.~\ref{Sec:Anomaly}. 
We examine the anomalies in detail by changing parameters
such as strength and type of the disorder, spatial dimensions,
and doping concentration.
We discuss the origin of these anomalies in Sec.~\ref{Sec:Origin}.
In Sec.~\ref{Sec:quantitative},
we examine quantitative aspects of the excitation spectrum
by applying the spectral function analysis
in Sec.~\ref{Sec:Analysis} to the numerical results.
We compare them with the analytical expressions
obtained in Sec.~\ref{Sec:Analytical}.

\subsection{Distribution form of disorder}
\label{Sec:Randomness}

In the following sections,
we consider the on-site disorder $\varepsilon_i = \delta\varepsilon$
and the bond disorder 
$t_{ij} = t + \delta t$ only for the nearest neighbor sites
where $\delta\varepsilon$ and $\delta t$ obey
one of the following three distribution functions.
The first one is the binary distribution in the form
\begin{equation}
P(x) = \frac12 \big[ \ 
\delta(\Delta) + \delta(-\Delta) \ \big],
\label{eq:binary}
\end{equation}
namely, $x = + \Delta$ or $-\Delta$ in the equal probability.
The second one is the Gaussian distribution in the form
\begin{equation}
P(x) = \frac{1}{\sqrt{2\pi} \Delta} 
\exp\Big(-\frac{x^2}{2 \Delta^2}\Big).
\label{eq:Gaussian}
\end{equation}
The last one is the box distribution in the form
\begin{equation}
P(x) = \frac{1}{2 p} 
\Theta(x+p) \big( 1 - \Theta(x-p) \big),
\label{eq:box}
\end{equation}
where $p = \sqrt{3} \Delta$ and
$\Theta(x)$ is the Heaviside step function.
The normalizations are taken 
to give the same second moment as
$\int x^2 P(x) dx = \Delta^2$.

In the following calculations, we consider the hypercubic lattice 
in $d$ dimensions, and we take the half-bandwidth $W=zt=2dt$ 
for $J_{\rm H} = \Delta = 0$ as an energy unit.
We change the value of $\Delta$ as a parameter
typically up to $0.5 W$.
It is difficult to determine the realistic value of $\Delta$ 
in low $T_{\rm C}$ manganites.
A rough estimate has been given by the first principle calculation,
which shows that $\Delta$ becomes the same order of magnitude 
of the half-bandwidth $W$.
\cite{Pickett1997}

\subsection{Anomaly in spin excitation spectrum}
\label{Sec:Anomaly}

\subsubsection{Dependence on strength and type of disorder}
\label{Sec:Randomness dependence}

We first show the change of the spin excitation spectrum
by controlling the strength of the disorder in Fig.~\ref{fig:de}.
Here, we fix the doping concentration at $x=0.3$,
where the hole density $x$ is defined as
$x = 1 - \langle \sum_{i}
c_{i}^\dagger c_{i} / N \rangle$ 
(we drop the spin index because the ground state is perfectly polarized).
Numerical calculations are performed 
for $N = 20 \times 20 \times 20$ site clusters
under the periodic boundary conditions.
The results are for the on-site disorder 
whose distribution is given by the binary form of Eq.~(\ref{eq:binary}).
The spectrum is obtained by
the random average of $A(\mib{q},\omega)$ 
in Eq.~(\ref{eq:Aqw}) over $16$ different realizations
of random configurations for each value of $\Delta$.
The gray-scale contrast shows the intensity of the spectrum.

\begin{figure}
\caption{
Spin excitation spectra at $x=0.3$ in the presence of the on-site disorder
in the binary distribution with (a) $\Delta = 0$,
(b) $\Delta = 0.1$, (c) $\Delta =0.2$, and (d) $\Delta = 0.3$, 
respectively.
}
\label{fig:de}
\end{figure}

In the case of $\Delta = 0$, i.e., 
in the absence of the disorder,
the spin excitation spectrum is given by a cosine dispersion
as shown in Fig.~\ref{fig:de} (a).
When we switch on the disorder, 
the excitation shows a finite linewidth
which becomes large as $\Delta$ increases.
Moreover, as clearly seen in Figs.~\ref{fig:de} 
(c) and (d), the spectrum shows some anomalies
in the large $q$ region.
There, we have additional broadening as well as
branching, that is, an emergence of an additional branch
which has significantly lower energy
than the original cosine like excitation.
For large values of $\Delta$,
the lower branch has a substantial weight
as shown in Fig.~\ref{fig:de} (d).
If one follows only the lower branch,
the spectrum appears to show softening
near the zone boundary.

These anomalous features are commonly observed
irrespective of the type of the disorder.
Figure~\ref{fig:distribution} shows the results
for different distribution functions; 
(a) is for the Gaussian distribution of Eq.~(\ref{eq:Gaussian}) and
(b) is for the box distribution of Eq.~(\ref{eq:box}).
In both cases, we have similar anomalous features
to those in Fig.~\ref{fig:de}. 

\begin{figure}
\caption{
Spin excitation spectra at $x=0.3$ in the presence of the on-site disorder
with $\Delta = 0.3$
in (a) the Gaussian distribution and (b) the box distribution.
}
\label{fig:distribution}
\end{figure}

We also examine the case of the bond disorder in Fig.~\ref{fig:bond}.
Here, we consider the binary distribution (\ref{eq:binary})
for $\delta t$.
In this case also, the spectrum shows additional broadening
which is more prominent and makes a branching obscure
compared to the cases of the on-site disorder 
in Figs.~\ref{fig:de} and \ref{fig:distribution}.
This is related to strong incoherence in the case of the bond disorder
which will be discussed quantitatively in Sec.~\ref{Sec:quantitative}.

\begin{figure}
\caption{
Spin excitation spectrum at $x=0.3$ in the presence of the bond disorder
in the binary distribution with $\Delta = 0.15$.
}
\label{fig:bond}
\end{figure}

\subsubsection{Spatial dimension dependence}
\label{Sec:Spatial dimension dependence}

Next, we examine systems in less spatial dimensions, namely,
in two and one dimensions.
In manganites, there are compounds which have strong
two-dimensional anisotropy, such as
the bilayer materials {\it A}$_3$Mn$_2$O$_7$ and
the single-layer materials {\it A}$_2$MnO$_4$.
Later, we will discuss the quantitative comparison
between our results and experimental ones in these compounds
in Sec.~\ref{Sec:Discussions}.
Here, we calculate $40 \times 40$ site clusters 
for $50$ different realizations of random configurations
in the two-dimensional (2D) case, and
$256$ site clusters
for $500$ configurations in the one-dimensional (1D) case.

\begin{figure}
\caption{
Spin excitation spectra in (a) two dimensions and (b) one dimension.
The results are for $x=0.3$ with the on-site disorder 
in the binary distribution with $\Delta=0.3$.
}
\label{fig:dimension}
\end{figure}

Figure~\ref{fig:dimension} shows the excitation spectra
in 2D and 1D systems at $x=0.3$ for the on-site disorder in
the binary distribution of Eq.~(\ref{eq:binary})
with $\Delta = 0.3$.
In both dimensions, the spectra show the anomalies clearly
as in three-dimensional (3D) systems
in Sec.~\ref{Sec:Randomness dependence}.
The important aspect is that
in lower dimensions the anomalous features appear more apparently.
Especially, in the 1D case in Fig.~\ref{fig:dimension} (b),
the detailed structure of the anomalies can be observed.
There, we have anticrossing with gap opening and
shadow-band like features.
From this systematic study for different dimensions, 
we speculate that the branching observed in higher dimensions
is a remnant of this anticrossing.

The anticrossing with the gap opening reminds us of 
the electron dispersion in systems
with spin density wave or collective phonon mode.
In these systems, magnons or phonons scatter the electrons
and open a gap with the anticrossing in the dispersion.
This analogy will be discussed further in Sec.~\ref{Sec:Friedel}.

\subsubsection{Doping dependence}
\label{Sec:Doping dependence}

Here, we study the doping dependence in a systematic way.
Figure~\ref{fig:doping3D} shows the results 
for different hole concentrations $x$ in 3D cases.
We consider here the on-site disorder whose distribution
is given by the binary form of Eq.~(\ref{eq:binary}) 
with $\Delta = 0.2$.
Figure~\ref{fig:doping3D} shows that 
the anomalous features become apparent as $x$ decreases.
This is probably because the disorder strength becomes
relatively large compared to the kinetic energy of electrons
which becomes small as $x$ decreases.
Moreover, we note that the positions of the anomalies appear to shift as $x$.

\begin{figure}
\caption{
Spin excitation spectra in three dimensions 
for (a) $x=0.5$, (b) $x=0.4$, (c) $x=0.3$, and
(d) $x=0.2$, respectively.
The results are for the on-site disorder 
in the binary distribution with $\Delta=0.2$.
}
\label{fig:doping3D}
\end{figure}

In order to observe the $x$ dependence more clearly,
we study the 1D systems under the same conditions.
Figure~\ref{fig:doping1D} shows the results.
As indicated in the figures, 
the position of the anomalous features clearly shift as the change of $x$, 
and is around the Fermi wavenumber $k_{\rm F}$
which is given by $\pi(1-x)$ in this 1D case
($k_{\rm F}$ is for the spinless fermion
in the perfectly polarized ground state).
We note also that the relatively weak anomalies can be
observed additionally at $q \sim \pi-k_{\rm F}, 2k_{\rm F}, 2(\pi-k_{\rm F})$, 
and so on.
These indicate that the anomalies are closely related to 
the fermionic degrees of freedom.
This observation provides a key to understand
the origin of the anomalies.

\begin{figure}
\caption{
Spin excitation spectra in one dimension
for (a) $x=0.5$, (b) $x=0.4$, (c) $x=0.3$, and
(d) $x=0.2$, respectively.
The results are for the on-site disorder 
in the binary distribution with $\Delta=0.2$.
The dashed lines represent the Fermi wavenumber at each $x$.
}
\label{fig:doping1D}
\end{figure}

\subsection{Origin of the anomaly}
\label{Sec:Origin}

\subsubsection{Friedel oscillation}
\label{Sec:Friedel}

We summarize the aspects of the anomalies of the spin excitation spectrum 
which are observed in the previous section.
(i) There appear additional broadening, branching,
anticrossing with gap opening, and shadow band.
(ii) These anomalous features are observed universally
for different types of the disorder, spatial dimensions and
doping concentrations.
(iii) The anomalies become more apparent in lower dimensions.
(iv) They appear most conspicuously at $\mib{q} \sim \mib{k}_{\rm F}$
while the additional anomalies can be seen
at $\mib{q} \sim \mib{Q}-\mib{k}_{\rm F}, 2\mib{k}_{\rm F}, 
2(\mib{Q}-\mib{k}_{\rm F})$, etc 
where $\mib{Q} = (\pi,\pi,\cdot \cdot \cdot)$.

As mentioned in Sec.~\ref{Sec:Spatial dimension dependence},
the anticrossing structure suggests 
a scattering mechanism for magnons.
Moreover, as shown in Sec.~\ref{Sec:Doping dependence},
this scattering is characterized by the Fermi wavenumber $\mib{k}_{\rm F}$,
which suggests that the scatterers are closely related 
with the fermionic degrees of freedom.
From these points, we consider that
the origin of the anomalies is the Friedel oscillation.
\cite{Friedel1953}

The mechanism is as follows.
In the presence of disorder, electrons tend to 
screen the effects of disorder, which leads to
the charge density correlation
with the wavenumber $2 \mib{k}_{\rm F}$.
This is the so-called Friedel oscillation.
\cite{Friedel1953}
In this DE system,
the charge density wave is equivalent to the spin density wave
since the ground state is the fully polarized
ferromagnetic state (half-metallic state
\cite{Park1998}).
Thus, we have the $2 \mib{k}_{\rm F}$ spin density correlation
due to the disorder though it is nonmagnetic. 
The spin density oscillation couples to magnon excitations and
scatters them.
This leads to the anticrossing features near $\mib{q} = \mib{k}_{\rm F}$.
(Note that the scattering wavenumber $2 \mib{k}_{\rm F}$ 
corresponds to the scattering 
between $\mib{q} \sim \mib{k}_{\rm F}$ and $-\mib{k}_{\rm F}$.)
The Friedel oscillation itself is a general phenomenon
in fermionic systems with disorder, and
its effect becomes relatively weaker in higher dimensions.
These aspects are consistent with our observations mentioned above.
Therefore, we conclude that the Friedel oscillation 
is the origin of the spin excitation anomalies in this system.

\subsubsection{Single impurity problem}
\label{Sec:Single impurity}

In order to confirm the picture based on the Friedel oscillation,
we examine the problem with a single impurity.
In this case, the charge density correlation is 
clearly found around the impurity position.
Figure~\ref{fig:single imp} (a) shows 
the density distribution in the real space 
when we introduce a single impurity
in the 1D system with $N = 256$.
Here, $n_i = \langle c_i^\dagger c_i \rangle$ and
we put the on-site potential at $i=128$
with relatively large $\varepsilon_i = -5$
to enlarge the impurity effect in the spectrum.
The electron density oscillates around the impurity site, 
which results in the singularity at $q = 2 k_{\rm F}$
in the density correlation function
as shown in Fig.~\ref{fig:single imp} (b).
Note that $q = 0.6\pi$ corresponds to 
$2k_{\rm F}$ with the hole Fermi wavenumber 
$k_{\rm F} = 0.3\pi$ 
(or $2\pi - 2k_{\rm F}$ with the electron Fermi wavenumber 
$k_{\rm F} = 0.7\pi$). 
Here, we define the density correlation function as
\begin{equation}
N(q) = \frac1N \sum_{ij} n_i n_j \
{\rm e}^{{\rm i} q (i-j)}.
\end{equation}
In this case of the single impurity also,
as shown in Fig.~\ref{fig:single imp} (c),
the spin excitation spectrum shows weak but obvious
anomalies at $k = k_{\rm F}, \pi-k_{\rm F}$, etc.,
which are similar to the results 
in the previous section \ref{Sec:Anomaly}.
The position of the anomalies is solely determined by $k_{\rm F}$
while their strength depends on the number and the strength 
of the disorder.
This strongly supports our scenario of the Friedel oscillation.

\begin{figure}
\caption{
(a) Local electron density  
when the on-site potential is finite only at the site $i=128$.
(b) Density correlation function.
(c) Spin excitation spectrum.
The dashed line in (c) represents the Fermi wavenumber.
}
\label{fig:single imp}
\end{figure}

\subsubsection{Comparison with random exchange Heisenberg model}
\label{Sec:Random Heisenberg}

In Sec.~\ref{Sec:Heisenberg}, we discussed 
the correspondence between the DE model and
the Heisenberg spin model within the spin wave approximation.
There, Eq.~(\ref{eq:J_ij}) gives a relation between these two models,
where the exchange coupling $J_{ij}$ is given by
the expectation value $\langle c_i^\dagger c_j \rangle$.
In order to show the fermionic effect in this quantity explicitly,
we study here the spin excitation
in the Heisenberg model with completely random exchange couplings.
Namely, we consider the ferromagnetic Heisenberg model with
the nearest-neighbor exchange couplings $J_{ij}^{({\rm r})}$ which take random
values from bond to bond and no longer satisfy Eq.~(\ref{eq:J_ij}).
This will elucidate a contrast from the random DE model (\ref{eq:H_DE}).
Within the lowest order of $1/S$ expansion,
this model leads to the effective spin-wave Hamiltonian in the form
\begin{equation}
{\cal H}_{i \neq j} = -2S J_{ij}^{({\rm r})},
\quad \
{\cal H}_{ii} = 2S \sum_{j} J_{ij}^{({\rm r})}.
\label{eq:H_Heis_eff}
\end{equation}
(See Appendix~\ref{AppendixA}.)
Although this effective Hamiltonian has a similar structure to
the magnon self-energy and satisfies the sum rule Eq.~(\ref{eq:sum rule}),
it is nothing to do with the fermionic degrees of freedom
in the DE model.
We calculate the spin excitation from Eqs.~(\ref{eq:H_Heis_eff}) in 1D. 
A typical spectrum is shown in Fig.~\ref{fig:Heis}.
Here, we choose $J_{ij}^{({\rm r})}$ randomly
from the range of
$1-\Delta \le J_{ij}^{({\rm r})} \le 1+\Delta$
with $\Delta = 0.25$.
We find no notable anomalies except for the broadening for the entire spectrum 
in contrast to the results in Sec.~\ref{Sec:Anomaly}.
This result supports the above picture and elucidates 
the relevance of the fermionic degrees of freedom of itinerant electrons
to the spin excitation in the random DE model.

\begin{figure}
\caption{
Spin excitation spectrum in the Heisenberg spin model
with random exchange couplings.
See the text for details.
}
\label{fig:Heis}
\end{figure}

\subsection{Quantitative aspects of the spectrum}
\label{Sec:quantitative}

As shown in Sec.~\ref{Sec:Anomaly},
although the spin excitation spectrum exhibits
significant anomalies near the Fermi wavenumber $\mib{k}_{\rm F}$,
it appears to be always single-peaked 
near the zone center $\mib{q} = 0$.
This allows us to apply the spectral function analysis
in Sec.~\ref{Sec:Analysis} 
in this regime.
In this section, we discuss the quantitative aspects
of the excitation spectrum by applying the spectral function analysis
to the numerical results 
in comparison with the analytical expressions
obtained in Secs.~\ref{Sec:Excitation energy} and
\ref{Sec:Linewidth}.
Throughout this section,
we consider the binary distribution (\ref{eq:binary})
for both cases of the bond disorder and the on-site disorder.

\subsubsection{Spin wave stiffness and magnon bandwidth}
\label{Sec:Stiffness}

\begin{figure}
\includegraphics[width=8cm]{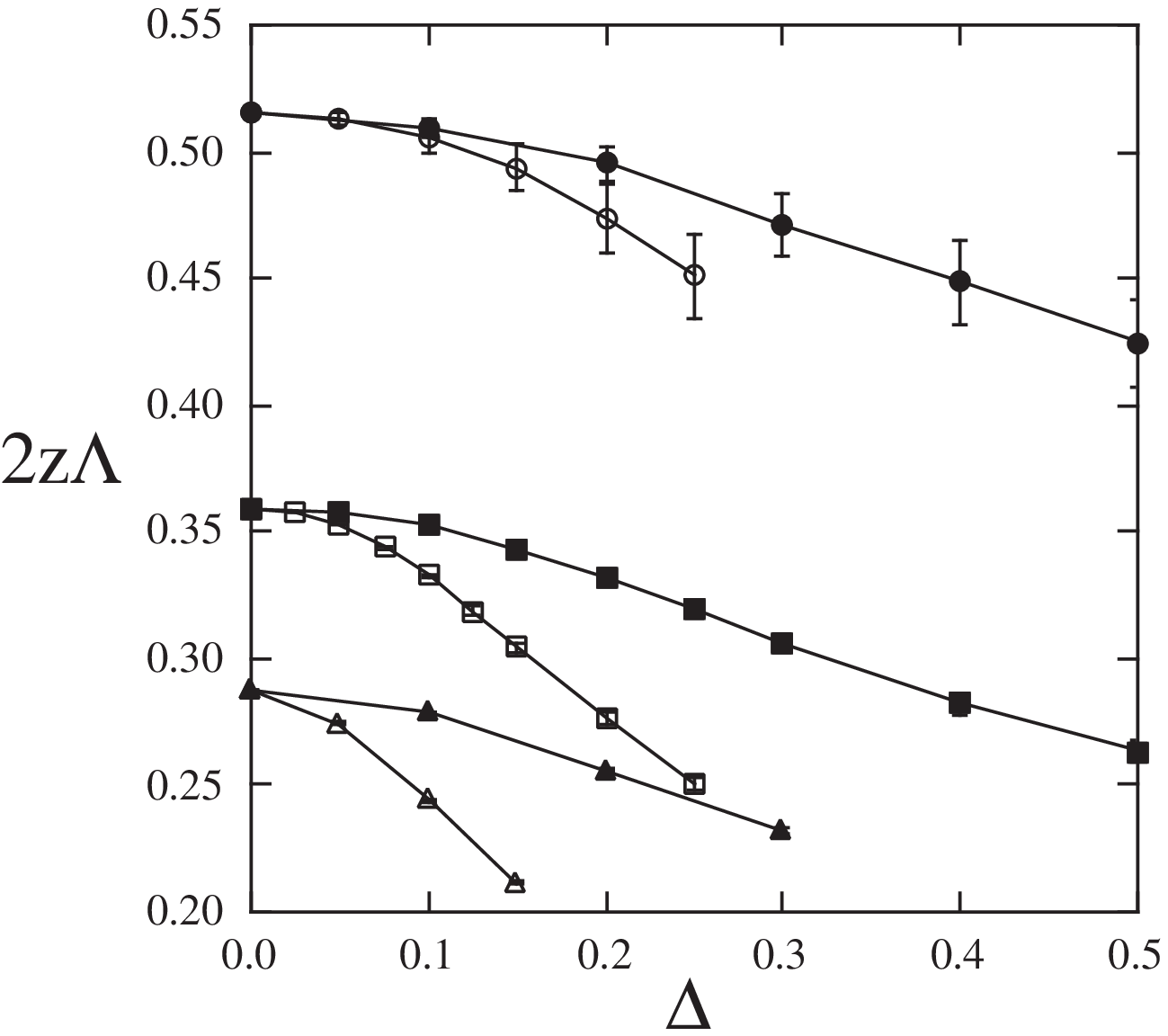}
\caption{
Spin stiffness at $x=0.3$ as a function of $\Delta$.
Circles, squares, and triangles represent the results 
in one, two, and three dimensions, respectively.
Open and filled symbols correspond to
the cases of the bond disorder and the on-site disorder, respectively.
The lines are guides for the eye.
}
\label{fig:stiffness}
\end{figure}

As shown in Eq.~(\ref{eq:Lambda to Ekin}),
the stiffness $\Lambda$ is proportional to 
the kinetic energy of electrons per site in our analysis.
Figure~\ref{fig:stiffness} plots the stiffness 
as a function of $\Delta$ at $x=0.3$,
which is calculated from the magnon self-energy by using Eq.~(\ref{eq:Lambda}).
In all dimensions, the stiffness decreases
as the disorder becomes strong. 
The asymptotic dependence in the limit of $\Delta \rightarrow 0$ 
appears to be $\Lambda \propto \Delta^2$.

\begin{figure}
\includegraphics[width=7.5cm]{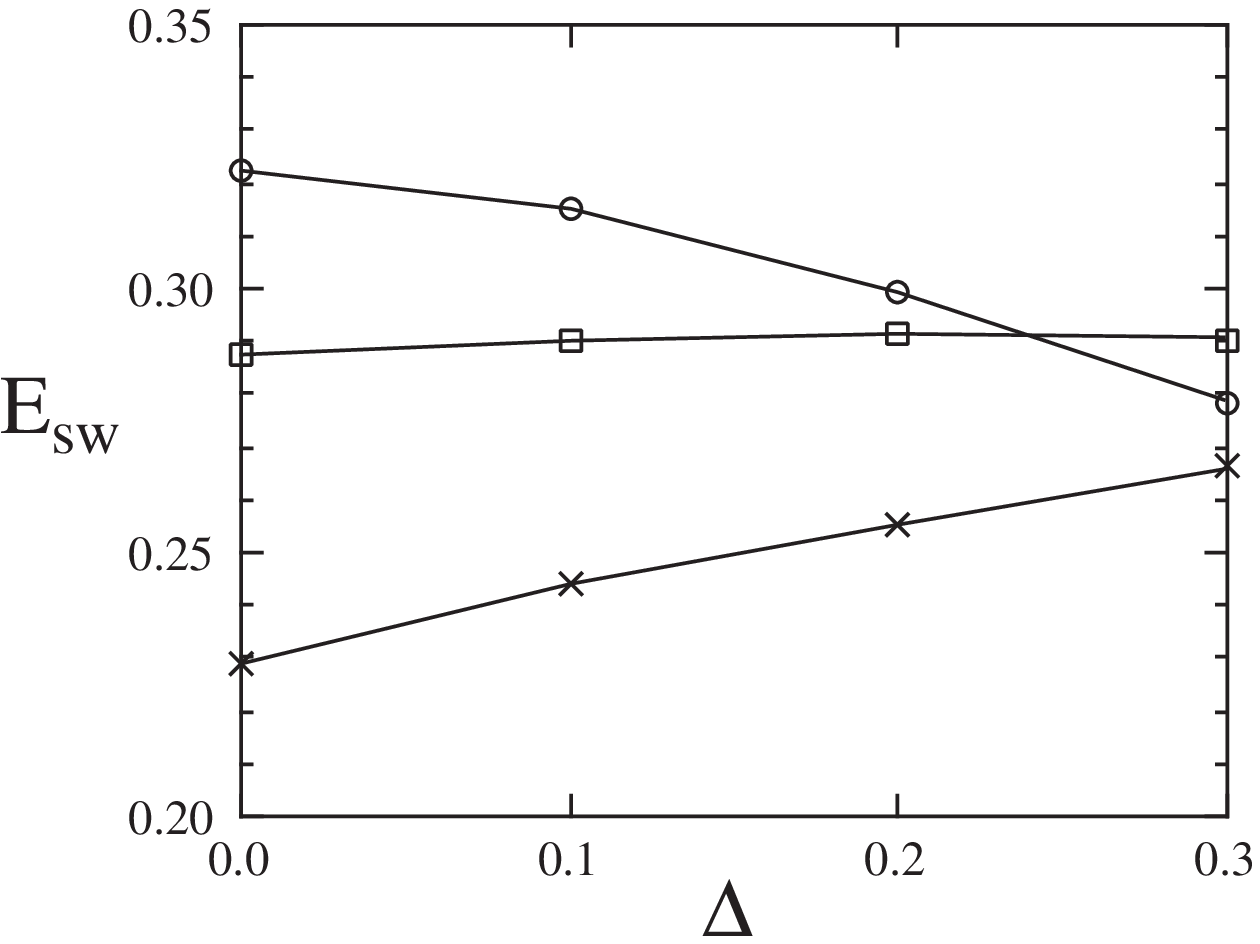}
\caption{
Total bandwidth of the spin excitation spectrum in three dimension 
with the on-site disorder.
Circles, squares, and crosses represent the results
for $x=0.4, 0.3$, and $0.2$, respectively.
The lines are guides for the eye.
}
\label{fig:E_sw vs de}
\end{figure}

Next we consider the bandwidth of the spin excitation.
In the pure system without disorder, the total bandwidth of
the spin wave excitation $E_{\rm sw}$ is 
determined by the spin stiffness $\Lambda$ in the form
$E_{\rm sw} = 2z \Lambda$.
\cite{Furukawa1996}
Thus, from Eq.~(\ref{eq:Lambda}), the spin wave bandwidth
$E_{\rm sw}$ is proportional to the kinetic energy of electrons
per site in the absence of disorder.
This relation no longer holds in the presence of disorder. 
We estimate the bandwidth $E_{\rm sw}$ 
by the peak energy of the highest excitation
at $\mib{q} = (\pi,\pi,\pi)$ in 3D systems.
Figure~\ref{fig:E_sw vs de} plots the estimates of $E_{\rm sw}$
as a function of the strength of the on-site disorder $\Delta$.
Although the stiffness always decreases as the disorder strength increases,
the change of the bandwidth $E_{\rm sw}$ 
depends on the hole concentration $x$:
$E_{\rm sw}$ decreases for $x \simge 0.3$, 
however, $E_{\rm sw}$ increases for $x \simle 0.3$.
At $x \sim 0.3$, $E_{\rm sw}$ remains almost constant in this range of $\Delta$.
This indicates that in our DE system at $x \sim 0.3$,
although the spectrum near the zone center becomes narrower 
due to the decrease of the stiffness as the disorder becomes stronger,
the total bandwidth of the spin excitation is almost unchanged.
Indeed, this behavior is observed in Fig.~\ref{fig:de}.

\subsubsection{Excitation energy and linewidth}
\label{Sec:omega and gamma}

\begin{figure}
\includegraphics[width=7cm]{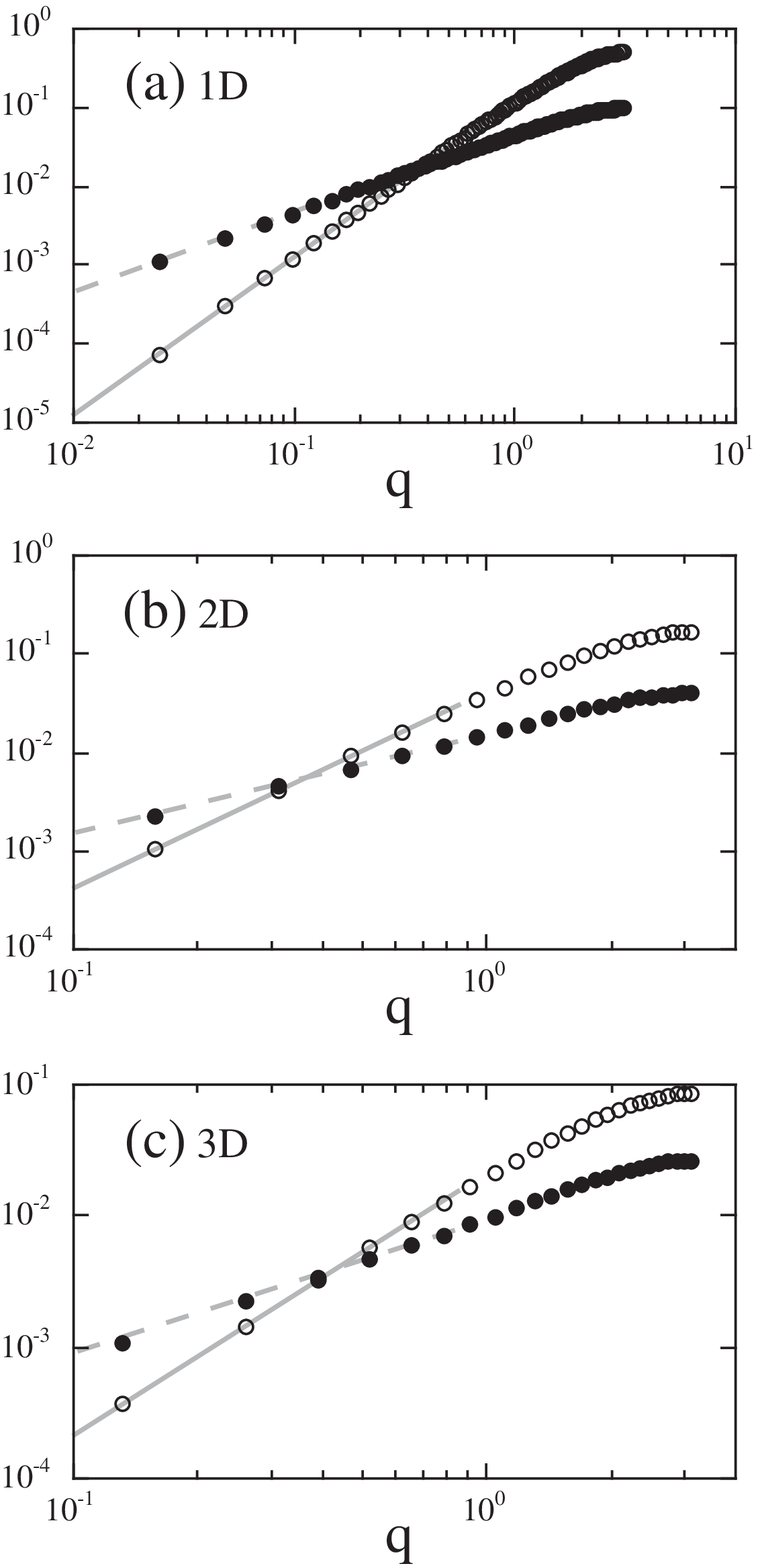}
\caption{
$q$ dependence of the excitation energy (open circles)
and the linewidth (filled circles) at $x=0.3$ 
in the presence of the on-site disorder 
in the binary distribution $\Delta=0.2$ 
in (a) one, (b) two, and (c) three dimensions, respectively, 
along $\mib{q} = (q,0,0,\cdot \cdot \cdot)$.
The gray solid (dashed) lines are the fits to $q^2$ ($q$).
}
\label{fig:q-dep of omega and gamma}
\end{figure}

The spectral function analysis in Sec.~\ref{Sec:Analysis} concludes 
the excitation energy $\omega_{\rm sw} \propto q^2$ and
the linewidth $\gamma \propto q$ 
in the limit of $q \rightarrow 0$.
Figure~\ref{fig:q-dep of omega and gamma} shows
the small $q$ behaviors of $\omega_{\rm sw}$ and $\gamma$
which are calculated from the numerical data
by using Eqs.~(\ref{eq:omega_sw}) and (\ref{eq:gamma}).
All the numerical results in different dimensions
show consistent $q$ dependences with the analytical predictions. 
We also confirm that there is the incoherent regime 
where $\gamma > \omega_{\rm sw}$ near $\mib{q}=0$.
These justify the applicability of 
the spectral function analysis which is based on 
the assumption of the single-peaked structure of the spectrum.

\begin{figure}
\includegraphics[width=8cm]{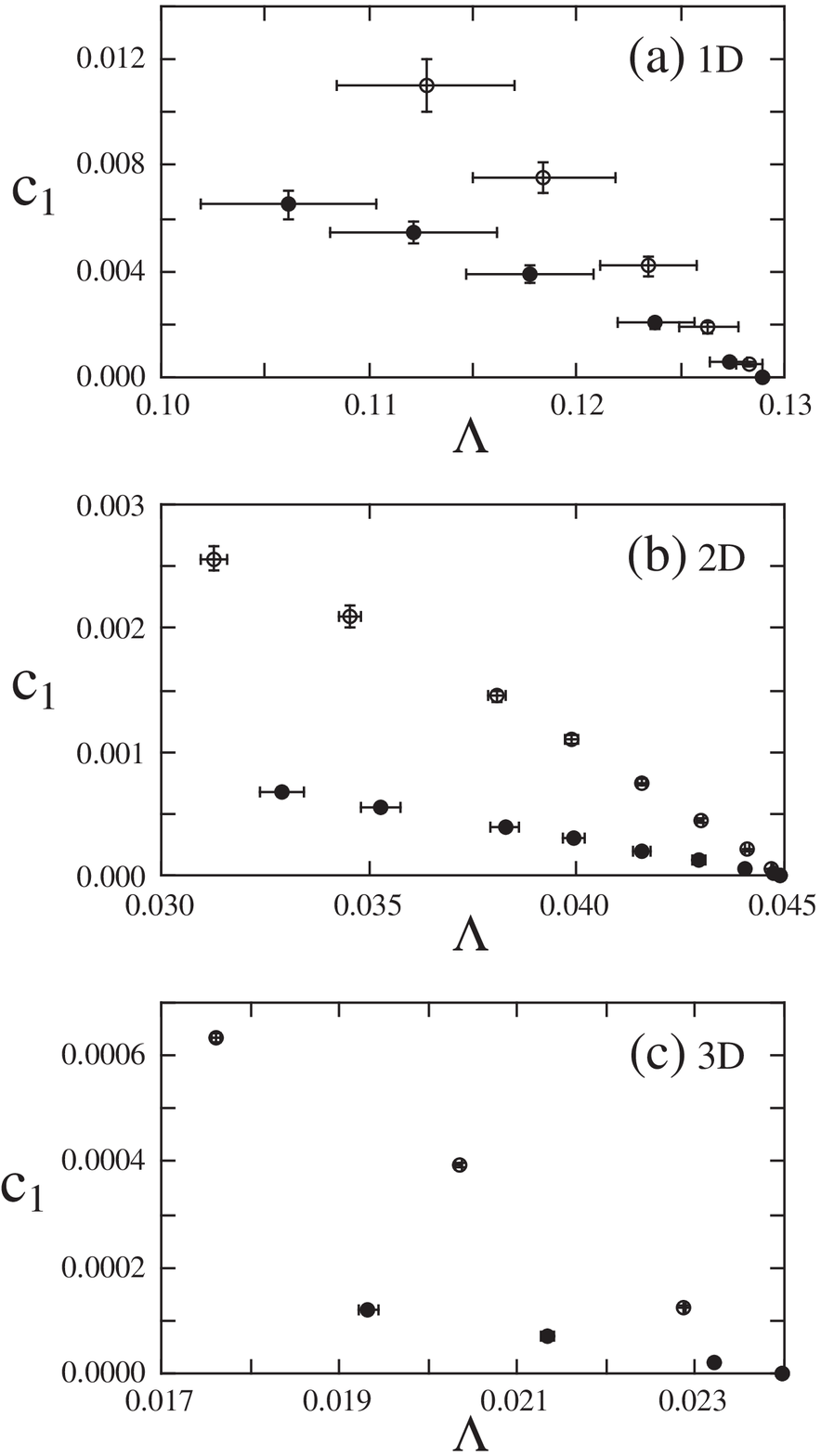}
\caption{
$q$ linear coefficient $c_1$ in the linewidth
as a function of the spin stiffness in
(a) one, (b) two, and (c) three dimensions, respectively.
Open and filled symbols correspond to
the cases of the bond disorder and the on-site disorder, respectively.
}
\label{fig:c1}
\end{figure}

Figure~\ref{fig:c1} plots the values of 
the coefficient of the $q$ linear term, 
$c_1$ in Eq.~(\ref{eq:gamma_q->0})
as a function of the stiffness $\Lambda$
by changing the disorder strength $\Delta$.
As $\Lambda$ decreases ($\Delta$ increases),
$c_1$ increases almost linearly to the decrease of $\Lambda$.
The interesting point is that
the coefficient $c_1$ increases much faster 
in the case of the bond disorder than the on-site disorder.
This indicates that the bond disorder induces 
larger $q$ linear term, namely, stronger incoherence 
in the magnon excitation than the on-site disorder.

\begin{figure}
\includegraphics[width=8cm]{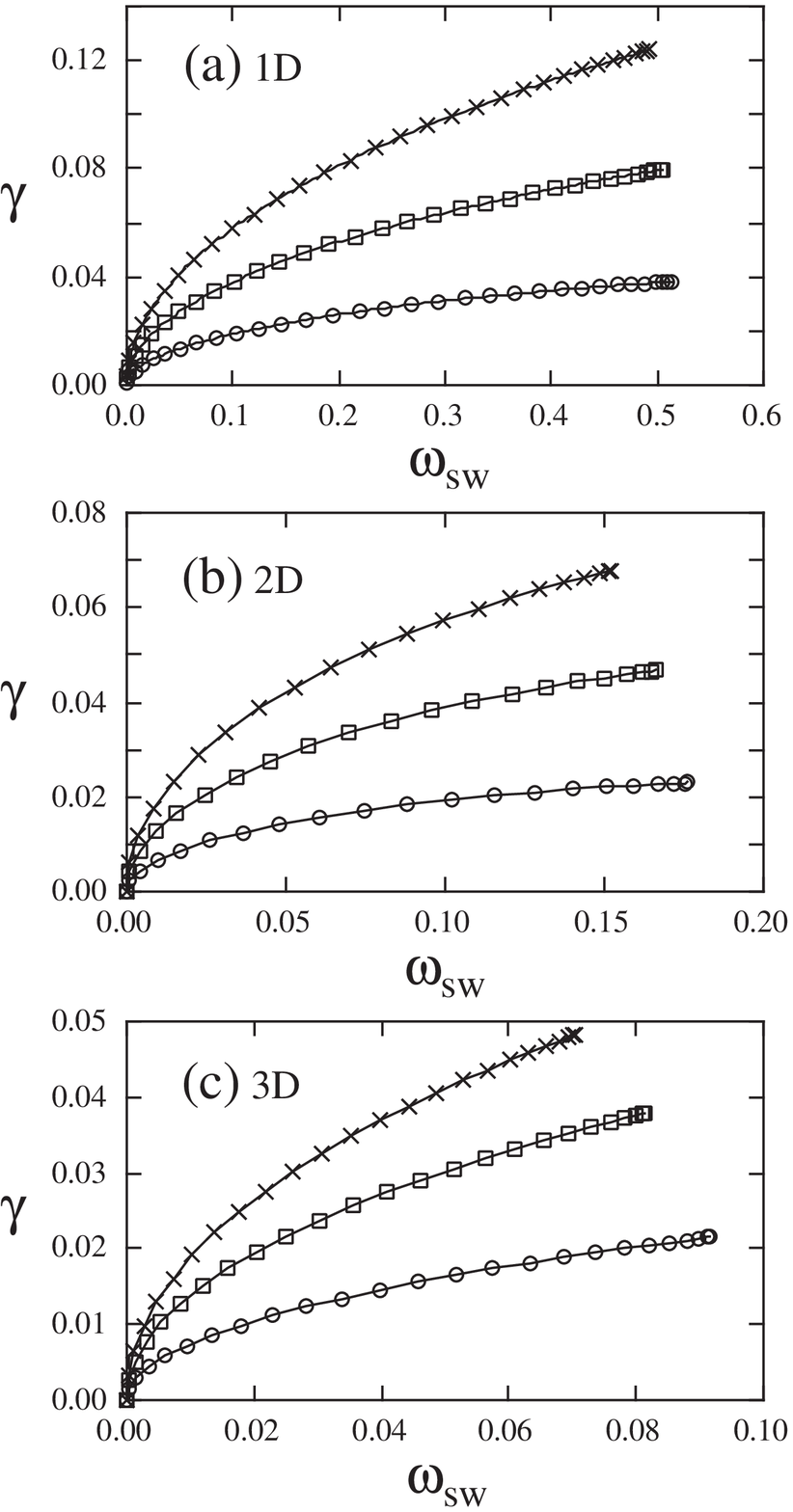}
\caption{
Linewidth as a function of excitation energy 
in the case of the bond disorder.
The results are for $x=0.3$ in 
(a) one, (b) two, and (c) three dimensions, respectively, 
along $\mib{q} = (q,0,0,\cdot \cdot \cdot)$.
Circles, squares, and crosses represent 
$\Delta = 0.05, 0.1$, and $0.15$, respectively.
}
\label{fig:gamma vs omega bond}
\end{figure}

\begin{figure}
\includegraphics[width=8cm]{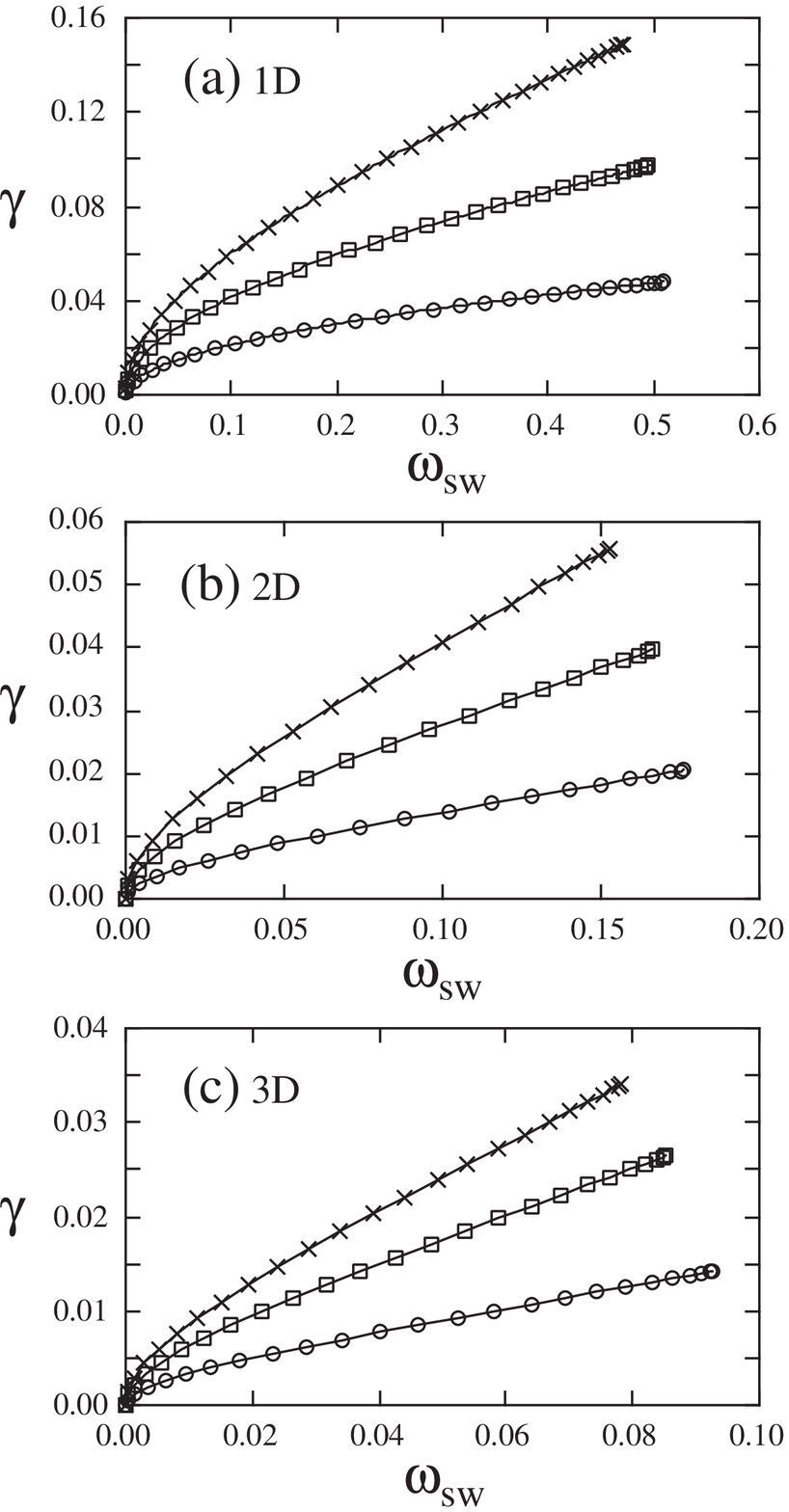}
\caption{
Linewidth as a function of excitation energy 
in the case of the on-site disorder.
The results are for $x=0.3$ in 
(a) one, (b) two, and (c) three dimensions, respectively, 
along $\mib{q} = (q,0,0,\cdot \cdot \cdot)$.
Circles, squares, and crosses represent 
$\Delta = 0.1, 0.2$, and $0.3$, respectively.
}
\label{fig:gamma vs omega onsite}
\end{figure}

This point is clearly demonstrated when we plot
the linewidth as a function of the excitation energy
as shown in Figs.~\ref{fig:gamma vs omega bond} and \ref{fig:gamma vs omega onsite}. 
For the bond disorder (Fig.~\ref{fig:gamma vs omega bond}),
the behavior $\gamma \propto \omega_{\rm sw}^{1/2}$
dominates the spectrum.
On the contrary, for the on-site disorder
(Fig.~\ref{fig:gamma vs omega onsite}),
the $\omega_{\rm sw}^{1/2}$ part is observed
only in the small $q$ regime, and
the behavior $\gamma \propto \omega_{\rm sw}$
is dominant in the wide region of $q$.
There, we have a crossover from the incoherent behavior
with $\gamma \propto \omega_{\rm sw}^{1/2} \propto q$
to the marginally coherent behavior
with $\gamma \propto \omega_{\rm sw} \propto q^2$.
Qualitative features are universal irrespective of the spatial dimensions.

We find here the qualitatively different behavior
between the bond disorder and the on-site disorder 
in the crossover from the incoherent to 
the marginally coherent regime.
Considering this difference and its origin,
we will discuss a general tendency of this crossover
and a realistic type of the disorder in real compounds
in Sec.~\ref{Sec:atomic or mesoscopic}.

\section{Discussions}
\label{Sec:Discussions}

\subsection{Spin excitation anomalies in low-$T_{\rm C}$ manganites}
\label{Sec:lowTc}

In high-$T_{\rm C}$ manganites, such as
La$_{1-x}$Sr$_x$MnO$_3$
and La$_{1-x}$Pb$_x$MnO$_3$ near $x=0.3$,
the spin excitation spectrum shows
a cosine like dispersion,
\cite{Perring1996,Moudden1998}
which can be well explained by the pure DE model
without disorder (Fig.~\ref{fig:de} (a)).
\cite{Furukawa1996}
Contrast to this canonical DE behavior,
low-$T_{\rm C}$ manganites such as
Pr$_{1-x}$Sr$_x$MnO$_3$,
La$_{1-x}$Ca$_x$MnO$_3$, 
and Nd$_{1-x}$Sr$_x$MnO$_3$
show the following anomalous features in the spin excitation spectrum.

One of the remarkable anomalies is called the softening.
Hwang {\it et al.} reported a significant softening of 
the spin wave dispersion near the zone boundary 
in Pr$_{1-x}$Sr$_x$MnO$_3$ at $x=0.37$.
\cite{Hwang1998}
Later, Dai {\it et al.} confirmed this result in several materials, 
and observed the optical phonon mode
in the energy range where the softening occurs.
\cite{Dai2000}
A similar spectrum with the optical phonon branch 
has been obtained also in the bilayer manganite, 
La$_{2-2x}$Sr$_{1+2x}$Mn$_2$O$_7$ at $x=0.4$.
\cite{Furukawa2000}
In this compound, however, 
a qualitatively different spectrum,
which is a cosine like form 
without any softening, has been observed 
in the same chemical formula but in a different sample crystal.
\cite{Perring2001}
This suggests the strong sample dependence of 
this softening behavior.

Another anomaly is the broadening. 
\cite{Hwang1998,Vasiliu-Doloc1998,Dai2000}
In compounds which show the softening, 
the linewidth of the excitation becomes significantly large
in the region where the softening is observed.
An important aspect is that
the linewidth remains finite even at the lowest temperature
where the ferromagnetic moment is fully saturated.
This point will be discussed in Sec.~\ref{Sec:Broadening}.

In compounds with relatively small $x$ (low doping),
the spin excitation shows another anomaly
such as the gap opening and/or the anticrossing.
\cite{Vasiliu-Doloc1998,Biotteau2001}
Note that the ground state in these compounds is 
ferromagnetic insulator and not metal.
The phase diagram in the small $x$ region shows
complicated transitions in structural and magnetic
degrees of freedom accompanied by a dimensional crossover.
\cite{Wollan1955,deGennes1960,Kawano1996a,Kawano1996b}

As shown in the results in Sec.~\ref{Sec:Results},
our simple model (\ref{eq:H_DE}), i.e., the DE model with disorder,
reproduces all the above anomalies,
at least, in a quantitative level. 
The softening appears in the lower energy branch in the anticrossing.
The additional broadening is observed 
in the large $q$ regime.
These anomalies become more conspicuous in lower doped case
and in lower dimensions.
These agreements suggest that the disorder is
a candidate to give a reasonable description on
the large spectral changes from the high-$T_{\rm C}$ to
the low-$T_{\rm C}$ manganites.
However, besides our disorder scenario,
there have been many other theoretical proposals
for the present problem.
We compare our results with those by other scenarios
in the following.

\subsection{Comparison with other theoretical results}
\label{Sec:Comparison theory}

\subsubsection{Softening}
\label{Sec:Softening}

Several mechanisms have been proposed 
for the softening near the zone boundary.
One is the purely magnetic origin, i.e.,
due to competitions 
between the DE ferromagnetic and
the superexchange antiferromagnetic interactions between localized spins
including longer ranged couplings.
\cite{Solovyev1999}
Other scenarios take account of a coupling
to other degrees of freedom.
One proposal is the coupling to orbital fluctuations,
in which orbital degrees of freedom in the doubly degenerate
$e_g$ orbitals are explicitly taken into account.
\cite{Khaliullin2000}
Another proposal is the coupling to phonons
which leads to the anticrossing
between the phonon excitation and the magnon excitation.
\cite{Furukawa1999b}

Although all these scenarios including our disorder scenario can reproduce 
the lower `softened' energy excitation in experimental results, 
there are some qualitative differences. 
In our results, the softening does not occur in the strict sense,
because the lower energy excitation 
appears just as a result of the anticrossing.
We still have the higher energy branch 
near the original dispersion in the absence of the disorder.
If one follows only the lower branch,
it looks like the softening as pointed out 
in Sec.~\ref{Sec:Randomness dependence}.
On the contrary, the mechanisms of the magnetic origin and 
of the magnon-orbital coupling origin
cause the softening of the original dispersion itself,
hence there is no higher energy branch.
In the scenario of the magnon-phonon coupling,
the spectrum shows a similar structure to our anticrossing.
However, in this case, the lower excitation near the zone boundary is
originally the optical phonon mode,
which extends over the whole Brillouin zone
with almost flat dispersion (optical mode).
There should be a finite energy excitation
even at the zone center additionally
to the gapless magnon excitation.
This is different from our results
in which the spectrum is single-peaked near the zone center.
These qualitative differences can be distinguished experimentally
by studying the high energy excitation 
in the wide region of the Brillouin zone.
This kind of experimental tests will be discussed and summarized
in Sec.~\ref{Sec:Exp tests}.

Besides the above qualitative differences,
our disorder scenario is different from 
the other proposals quantitatively.
In other scenarios, it should be noted that
the relative importance of the additional couplings is determined by
the change of the DE interaction, namely, the bandwidth,
since the strength of them does not change so much 
by the {\it A}-site substitution.
In this sense, these scenarios attribute 
the origin of the change of the spectrum
from high-$T_{\rm C}$ to low-$T_{\rm C}$ compounds
to the change of the bandwidth.
As discussed previously, however,
the actual change of the bandwidth in these compounds
is very small.
Therefore, it appears to be difficult
for these scenarios to explain
the large changes of the spin excitations
in a quantitative manner.
On the contrary, 
the strength of the disorder can be largely
affected by the {\it A}-site substitution
even if the change of the averaged lattice structure is small.
This appears to favor our disorder scenario.

\subsubsection{Broadening}
\label{Sec:Broadening}

As discussed in Sec.~\ref{Sec:lowTc},
the experimental results show 
the intrinsic linewidth at the lowest temperature.
This suggests that the spin wave excitation is no longer
the eigenstate of the system at the ground state. 
We discuss here two different mechanisms for 
this zero-temperature broadening.
One is effective even in the absence of disorder and
the other is our disorder scenario.

Even in the pure DE model without disorder,
the spin wave is not the exact eigenstate of the system.
\cite{Kaplan1997}
The coupling with charge degrees of freedom leads
to a finite lifetime for the spin wave excitations.
This zero-temperature broadening in the pure DE model
has been discussed by the spin wave approximation.
\cite{Golosov2000,Shannon2002}
In the lowest order of the $1/S$ expansion,
the DE model can be mapped to the pure Heisenberg model
as discussed in Appendix~\ref{AppendixA}.
Hence, the spin wave is the exact eigenstate and
the linewidth is zero in $O(1/S)$.
\cite{Furukawa1996}
In the higher order of $1/S$, however, 
we have a finite imaginary part in the magnon self-energy, 
which leads to a linewidth in the spin excitation spectrum.
Up to $O(1/S^2)$, this linewidth is predicted to show 
the $q$ dependence as $\gamma \propto q^{d+3}$ 
in the limit of $q \rightarrow 0$,
where $d$ is the spatial dimension of the system.
Thus, the linewidth is predicted to behave as
$\gamma \propto q^6$ in 3D and
$\gamma \propto q^5$ in 2D systems.

In the presence of disorder,
our results indicate that the spin excitation spectrum shows
a finite linewidth even in the lowest order of $1/S$ expansion.
As discussed in Sec.~\ref{Sec:self-energy},
this is an inhomogeneous broadening and
distinguished from the above broadening due to 
the finite lifetime of quasiparticle excitations.
In our results, the linewidth shows the $q$ linear behavior
in the limit of $q \rightarrow 0$, and
crossovers to the $q^2$ behavior as $q$ increases.
These behaviors do not depend on the spatial dimension $d$.
Therefore, it is possible to determine 
which scenario is relevant in real compounds
by studying the $q$ dependence of the linewidth 
in both 3D and 2D systems.

In 2D (bilayer) compounds, recently, detailed analyses have been done
on the $q$ dependence of the linewidth.
\cite{Perring2001}
The marginally coherent behavior
$\gamma \propto \omega_{\rm sw} \propto q^2$ is observed 
in the wide region of $\mib{q}$.
It is difficult to explain this $q$ dependence
by the former mechanism 
which predicts $\gamma \propto q^5$ in 2D.
In our disorder scenario, the marginally coherent behavior
is actually found as shown in Sec.~\ref{Sec:omega and gamma}.
This supports our scenario.

Unfortunately, the linewidth in the small $q$ region
has not been fully investigated experimentally, especially
in 3D compounds in the ferromagnetic metallic regime.
\cite{Hwang1998,Dai2000}
Detailed information near the zone center
especially for $q \simle \pi/5$ is lacking.
Further experimental study is desired to examine
the asymptotic $q$ dependence.

In our results, the crossover between the incoherent regime and
the marginally coherent regime appears to depend on
the details of the disorder as mentioned 
in Sec.~\ref{Sec:omega and gamma}.
The dominance of the marginally coherent behavior 
in experimental results provides us a microscopic picture
on the disorder in real compounds.
This will be discussed in Sec.~\ref{Sec:atomic or mesoscopic}
in detail.

\subsubsection{Anticrossing and gap opening}
\label{Sec:Anticrossing}

As mentioned in Sec.~\ref{Sec:lowTc},
the gap opening and/or the anticrossing have been experimentally observed 
only in the low doped region ($x \simle 0.2$)
where $T_{\rm C}$ is largely reduced from the value at $x \simeq 0.3$. 
\cite{Vasiliu-Doloc1998,Biotteau2001}
In this region, the kinetic energy
becomes small partly because the number of carriers decreases.
There, the system is insulating with ferromagnetism or spin canting
as well as the orthorhombic lattice distortion.
\cite{Wollan1955,deGennes1960,Kawano1996a,Kawano1996b}
Hence, there should be competitions among several interactions
such as ferromagnetic DE, antiferromagnetic superexchange, 
and electron-lattice interaction.

We note that, in the 3D system, the values of $\Delta$ in the present calculations
are much smaller than the critical value for the Anderson localization, 
\cite{Li1997,Sheng1997}
hence, the ground state remains the ferromagnetic metal in our results.
In lower dimensions, of course, electrons are easily localized at small $\Delta$.
The ferromagnetic or the spin canting state indicates 
the relevance of the DE interaction even in the insulating states. 
The charge/spin oscillation due to the disorder is expected 
even in these insulators 
although the Fermi wavenumber is no longer well defined.
Hence, we expect similar anomalies in these insulating regions.

Besides our disorder scenario, we have at least 
two mechanisms which cause the anticrossing and/or the gap opening 
in the spin excitation spectrum. 
One is the superlattice magnetic structure such as the spin canting.
In this case, the spin excitation spectrum shows a gap at 
the wavenumber for the corresponding magnetic periodicity.
The spectrum consists of a single branch in the extended Brillouin zone and
deviates from the cosine like form only near the gap.
The small gap observed at $\mib{q} = (0,0,\pi/2)$
in La$_{0.85}$Sr$_{0.15}$MnO$_3$ appears to be explained by this scenario
since the corresponding superlattice peak is observed.
\cite{Vasiliu-Doloc1998}
This behavior should be distinguished from 
the anticrossing or the branching in our results.

The other mechanism is the magnon-phonon coupling 
as mentioned in Sec.~\ref{Sec:Softening}. 
\cite{Furukawa1999b}
In this scenario, the anticrossing with the gap opening 
appears at the wavenumber where the magnon dispersion crosses
the optical phonon branch. 
The excitation spectrum looks similar to our results.
Possible experimental tests to distinguish these scenarios
will be discussed in Sec.~\ref{Sec:Exp tests}.

\subsection{Atomic vs. mesoscopic scale disorder}
\label{Sec:atomic or mesoscopic}

Our results in Sec.~\ref{Sec:omega and gamma}
elucidate a difference in the spin excitation
between the bond disorder and the on-site disorder.
In the case of the on-site disorder,
the crossover from the incoherent behavior
$\gamma \propto \omega_{\rm sw}^{1/2} \propto q$
to the marginally coherent one 
$\gamma \propto \omega_{\rm sw} \propto q^2$
takes place in the relatively small $q$ region, and
the latter behavior is dominant in the wide region of the spectrum.
On the other hand, in the case of the bond disorder,
the incoherent behavior dominates over the whole $q$ region.
The bond disorder tends to make the spectrum
more incoherent than the on-site disorder in the present model.

Since the incoherence comes from 
the local fluctuations of $J_{ij}$ in Eq.~(\ref{eq:J_ij})
as discussed in Sec.~\ref{Sec:Incoherence},
this difference indicates that
the bond disorder leads to relatively 
larger fluctuations of $J_{ij}$
than the on-site disorder in the present model.
This is qualitatively understood as follows.
In Eq.~(\ref{eq:J_ij}), there are two sources of
the local fluctuations of $J_{ij}$.
One is the transfer integral $t_{ij}$ and
the other is the expectation value 
$\langle c_i^\dagger c_j \rangle$.
In the case of the on-site disorder, 
$t_{ij}$ is taken to be uniform for all the nearest neighbor pairs,
hence, the local fluctuations come only from 
$\langle c_i^\dagger c_j \rangle$.
This expectation value is disordered in the presence of disorder,
however, it has some correlations due to 
the Friedel oscillation as discussed in Sec.~\ref{Sec:Origin}.
On the contrary, in the case of the bond disorder,
$t_{ij}$ is taken to be completely independent
from bond to bond in our model.
This local fluctuation of $t_{ij}$ is directly reflected in
that of $J_{ij}$ in Eq.~(\ref{eq:J_ij}).
Even if there are some spatial correlations
in $\langle c_i^\dagger c_j \rangle$,
the multiplication of $t_{ij}$ makes $J_{ij}$ more disordered.
Thus, in the present model, we expect a larger local fluctuations
in the case of the bond disorder than the on-site disorder.

However, this does not mean that 
the bond disorder always makes
the spin excitation spectrum more incoherent
than the on-site disorder in general.
The stronger incoherence in the present results is
due to the specific form of $t_{ij}$ in our model.
Even in the case of the bond disorder, 
the incoherence is possibly suppressed 
if we have some spatial correlations in $t_{ij}$.
Instead, the important conclusion is that 
the atomic-scale disorder 
leads to more incoherent excitation
than the spatially-correlated or 
the mesoscopic-scale disorder.

As discussed in Sec.~\ref{Sec:Broadening},
the experimental results in the bilayer compounds
show the dominance of the marginally coherent regime.
From the above conclusion, 
this indicates that in real materials 
the mesoscopic-scale disorder
is more relevant than the atomic-scale disorder
if the disorder plays a major role
in the spin excitation spectrum.
Possible sources of the mesoscopic-scale disorder
are a large-scale inhomogeneity or clustering of the {\it A} ions,
twin lattice structure, and grain boundaries.

\subsection{Experimental tests}
\label{Sec:Exp tests}

We propose here some experiments to test our disorder scenario.

(i) Sample dependence:
Our scenario predicts a correlation between 
the sample quality and the anomalies in the spin excitation spectrum.
Even in the same chemical formula,
the extent of the anomalies may depend on the sample quality.
The purity of samples can be measured independently,
for instance, by the magnitude of the residual resistivity.
In this test, we note that
the {\it A}-site ordered manganites 
\cite{Millange1998,Nakajima2002,Akahoshi2003}
will be useful.
In these materials, the disorder strength due to the alloying effect 
in the {\it A}-site ions can be tuned 
to some extent by the careful treatment of the syntheses.
It is interesting to compare the spin excitation spectrum 
in the {\it A}-site ordered compound with that in 
the ordinary solid-solution ({\it A}-site disordered) compound
in the same chemical formula.

(ii) Comparison with the Fermi surface: 
The spin excitation anomalies in our results appear 
strongly near the Fermi wavenumber.
It is crucial to compare the position of the anomalies with 
the information of the Fermi surface. 
The Fermi surface can be independently determined,
for instance, by the angle-resolved photoemission spectroscopy.

(iii) Doping dependence:
Closely related to the above test (ii),
the doping dependence is also a key experiment.
In our scenario, by controlling the doping concentration,
the position of the anomalies shifts 
due to the change of the Fermi surface. 
Moreover, the anomalies become more conspicuous 
for lower doping concentration.

(iv) High energy excitation:
Our scenario predicts the anticrossing instead of the softening.
Hence, near the zone boundary, 
there remains the higher energy excitation above the lower energy branch 
which looks like to be softened.
As discussed in Sec.~\ref{Sec:Softening},
this point is different from 
several other theoretical proposals.
At the same time, it is important to identify
the optical phonon mode, 
especially near the zone center, 
for examining the relevance of the magnon-phonon coupling
as discussed in Sec.~\ref{Sec:Softening}.
Therefore, it is crucial to study the higher energy region 
over the whole Brillouin zone.

(v) $q$ dependence of the linewidth:
Our results indicate that irrespective of the spatial dimension,
the linewidth of the spectrum
is proportional to $q$ near the zone center
and shows a crossover to $q^2$ behavior as $q$ increases.
As discussed in Sec.~\ref{Sec:omega and gamma},
this $q$ dependence appears to be consistent with 
the experimental result in the bilayer manganite.
In all other scenarios, the linewidth in the ground state
comes from the interplay between charge and spin
which is discussed in Sec.~\ref{Sec:Broadening}.
This predicts $q^{d+3}$ behavior near the zone center.
It is strongly desired to examine the linewidth more quantitatively
especially in 3D compounds.

\section{Summary and concluding remarks}
\label{Sec:Summary}

In this paper, we have discussed the disorder effects 
on the spin excitation spectrum in the double exchange model.
We have applied the spin wave approximation 
in the lowest order of $1/S$ expansion.
The analytical expressions are obtained for
the spin stiffness, the excitation energy, and the linewidth. 
The most important result revealed 
by the extensive numerical calculations is that
the spin excitation spectrum shows 
some anomalies in the presence of the disorder,
such as broadening, branching, or
anticrossing with gap opening.
The origin of these anomalies is the Friedel oscillation:
Disorder causes the $2k_{\rm F}$ oscillation 
of the spin and charge density
in the fully polarized ferromagnetic state, which scatters
the magnons to cause the anticrossing in their dispersion.
Thus, the spin excitation anomalies are
the consequence of the strong interplay between spin and charge
degrees of freedom.
Our results have been compared with experimental results 
in the {\it A}-site substituted manganites 
which shows relatively low-$T_{\rm C}$, and
the agreement is satisfactory.
We have also compared our results with other theoretical proposals
for the anomalous spin excitation in these manganites, 
and clarified the advantages of our disorder scenario.

Another important finding is 
the incoherence of the magnon excitation. 
We have shown that near the zone center 
the linewidth is proportional to the magnitude of the wavenumber $q$,
which becomes larger than
the excitation energy which scales to $q^2$.
This incoherence comes from the local fluctuations
of the transfer energy of itinerant electrons.
As $q$ increases, we have obtained the crossover
from this incoherent behavior 
to the marginally coherent one in which
both the linewidth and the excitation energy are
proportional to $q^2$.
We found that this crossover behavior depends on the nature of the disorder.
Comparing with experimental results in which
the marginally coherent behavior is dominant,
we have concluded that in real materials
the spatially-correlated or mesoscopic-scale disorder is relevant
compared to the local or atomic-scale disorder.

All the above anomalous features are obtained 
in the lowest order of the $1/S$ expansion, $O(1/S)$.
Up to $O(1/S)$, we have shown that the double exchange model
with disorder is effectively mapped to the Heisenberg spin model
with random exchange couplings. 
In the absence of the disorder, the corresponding Heisenberg model 
has the uniform nearest neighbor exchanges, 
whose spin excitation spectrum shows the cosine dispersion.
\cite{Furukawa1996,FurukawaPREPRINT}
Even in this pure case, in the higher order of $1/S$, 
there are some deviations from this cosine dispersion
such as the magnon bandwidth narrowing or the softening, and the broadening.
\cite{Kaplan1997,Golosov2000,Shannon2002}
However, these higher order corrections are known to be small 
near the hole concentration $x = 0.3$ where we are interested in.
\cite{Kaplan1997}
Moreover, the spin excitation anomalies we obtained here are 
the primary effects in the $1/S$ expansion 
compared to the higher-order corrections.
Therefore, we believe that 
our results are relevant to understand the anomalous spin excitations 
in real compounds.
Our results give a comprehensive understanding of 
the systematic changes of the spin excitation spectrum
in the {\it A}-site substituted manganites.
Moreover, the disorder explains well 
the experimental facts which cannot be understood 
only by the bandwidth control, 
such as the rapid decrease of $T_{\rm C}$. 
\cite{MotomePREPRINT}
Therefore, we conclude that
the disorder plays an important role 
in the {\it A}-site substitution of the ionic radius control
in CMR manganites.

We comment on the relevance of another additional couplings
due to the influence of the multicritical phenomena.
As decreasing the ionic size of the {\it A}-site atoms, 
finally the system becomes the charge-ordered insulator
with the orbital and lattice orderings, 
the ferromagnetic insulator, the antiferromagnetic insulator, or 
the spin glass insulator. 
\cite{Imada1998,Hwang1995,Goodenough1955,DeTeresa1997,Tomioka2002}
The phase diagram often shows the multicritical behavior. 
These phase transitions might not be explained by the disorder alone.
In the close vicinity of the multicritical phase boundary, 
we expect large fluctuations and critical enhancement
of additional couplings, 
which possibly accelerate the suppression of $T_{\rm C}$ and
enhance the anomalies in the spin excitation spectrum. 
However, the multicritical phenomena is essentially
the first order phase transition.
Hence, far from the phase boundary, for instance, in LCMO compounds, 
we believe that these additional couplings becomes irrelevant and
the disorder plays a major role.

\section*{Acknowledgment}

Y. M. acknowledges N. Nagaosa, Y. Tokura, and I. Solovyev
for suggestive discussions.
This work is supported by  ``a Grant-in-Aid from
the Mombukagakushou''.

\appendix

\section{}
\label{AppendixA}

In this Appendix, we show 
that the magnon self-energy (\ref{eq:Pi_ij_by_B_ij}) is equivalent to
the effective spin-wave Hamiltonian for the Heisenberg spin model
in the lowest order of the $1/S$ expansion.
\cite{FurukawaPREPRINT}

We begin with model~(\ref{eq:H_DE}) 
in the limit of $J_{\rm H}/t \rightarrow \infty$.
In this limit, 
since the conduction electron spin $\mib{\sigma}$
is completely parallel to
the localized spin  at each site,
states with $\mib{\sigma}$ antiparallel to
$\mib{S}$ are projected out.
Then, we have the effective spinless-fermion model as
\cite{Anderson1955}
\begin{equation}
{\cal H} = -\sum_{ij} (  \tilde{t}_{ij} 
c_i^\dagger c_j + {\rm h.c.}  ) + 
\sum_i \varepsilon_i c_i^\dagger c_i,
\label{eq:H_DE_limit}
\end{equation}
where the transfer integral $\tilde{t}_{ij}$ depends on
the relative angle of localized spins as
\begin{equation}
\tilde{t}_{ij} = t_{ij}
\Big[ \cos\frac{\theta_i}{2} \cos\frac{\theta_j}{2} + 
\sin\frac{\theta_i}{2} \sin\frac{\theta_j}{2}
{\rm e}^{- {\rm i} (\phi_i - \phi_j)}
\Big],
\end{equation}
with $S_i^x = S \sin\theta_i \cos\phi_i$,
$S_i^y = S \sin\theta_i \sin\theta_i$, and 
$S_i^z = S \cos\theta_i$.
Hereafter, we treat the localized spins $\mib{S}$
as classical objects.
The transfer integral in Eq.~(\ref{eq:H_DE_limit}) is a complex variable.
The real and imaginary parts are given by
\begin{eqnarray}
{\rm Re} \ \tilde{t}_{ij} &=& 
t_{ij} \sqrt{\frac12 \Big( 1 + \frac{\mib{S}_i \cdot \mib{S}_j}{S^2}\Big)},
\label{eq:Re t_ij}
\\
{\rm Im} \ \tilde{t}_{ij} &=&
\frac{t_{ij}}{2} \sqrt{\frac{S^2}{(S+S_i^z)(S+S_j^z)}}
\frac{S_i^y S_j^x - S_i^x S_j^y}{S^2}, 
\label{eq:Im t_ij}
\end{eqnarray}
respectively.
Note that ${\rm Im} \ \tilde{t}_{ij}$ is antisymmetric
while ${\rm Re} \ \tilde{t}_{ij}$ is symmetric,
as expected from $\tilde{t}_{ij} = \tilde{t}_{ji}^*$.

Now we apply the spin wave approximation to model~(\ref{eq:H_DE_limit}).
We use here the Holstein-Primakoff transformation
\begin{eqnarray}
S_i^z &=& S - a_i^\dagger a_i,
\nonumber \\
S_i^x &\simeq& \sqrt{\frac{S}{2}} (a_i^\dagger + a_i), \ \ 
S_i^y \simeq {\rm i} \sqrt{\frac{S}{2}} (a_i^\dagger - a_i), \ 
\label{eq:HPtrsf}
\end{eqnarray}
and take account of the $1/S$ expansion up to the leading order of $O(1/S)$.
Substituting Eqs.~(\ref{eq:HPtrsf}) 
into Eqs.~(\ref{eq:Re t_ij}) and (\ref{eq:Im t_ij}),
we obtain
\begin{eqnarray}
\frac{{\rm Re} \ \tilde{t}_{ij}}{t_{ij}} &\simeq& 
1 + \frac{1}{4S}
(a_i^\dagger a_j + a_j^\dagger a_i 
- a_i^\dagger a_i - a_j^\dagger a_j),
\\
\frac{{\rm Im} \ \tilde{t}_{ij}}{t_{ij}} &\simeq&
\frac{1}{4S} (a_i^\dagger a_j - a_j^\dagger a_i).
\end{eqnarray}
Hence, the effective magnon-electron
Hamiltonian up to $O(1/S)$ is given by
\begin{eqnarray}
{\cal H} &\simeq& -\sum_{ij} t_{ij}
\Big[ 1 + \frac{1}{4S} (a_i^\dagger a_j + a_j^\dagger a_i 
- a_i^\dagger a_i - a_j^\dagger a_j)
\Big]
\nonumber \\
&& \quad \quad \quad \quad \quad
\times (c_i^\dagger c_j + c_j^\dagger c_i)
\nonumber \\
&& + \frac{1}{4S} \sum_{ij} t_{ij}
(a_i^\dagger a_j - a_j^\dagger a_i)
(c_i^\dagger c_j - c_j^\dagger c_i)
\nonumber \\
&&+ \sum_i \varepsilon_i c_i^\dagger c_i.
\label{eq:H_DE_mag_ele}
\end{eqnarray}

Finally, we obtain the effective magnon Hamiltonian
by tracing out the fermion degrees of freedom 
in Eq.~(\ref{eq:H_DE_mag_ele}).
Up to $O(1/S)$, the trace can be calculated as
the expectation value for the ground state without any magnon.
The result is given by
\begin{equation}
{\cal H} \simeq -\sum_{ij} \frac{t_{ij}}{2S}
\langle c_i^\dagger c_j \rangle
(a_i^\dagger a_j + a_j^\dagger a_i 
- a_i^\dagger a_i - a_j^\dagger a_j),
\label{eq:H_DE_eff}
\end{equation}
up to irrelevant constants.
Here we use the general relation
$\langle c_i^\dagger c_j \rangle = \langle c_j^\dagger c_i \rangle$
for the perfectly polarized ground state without degeneracy.
Note that this effective Hamiltonian is derived
generally for both the bond disorder and the on-site disorder.
From the eigenvalues and the eigenvectors for Eq.~(\ref{eq:H_DE_eff}),
we obtain the spectral function $A(\mib{q},\omega)$.
Hence, comparing with Eqs.~(\ref{eq:eigenPi}) and (\ref{eq:Aqw}),
the static part of the magnon self-energy $\Pi(\omega=0)$ is
equivalent to the effective magnon Hamiltonian~(\ref{eq:H_DE_eff}).

This effective Hamiltonian (\ref{eq:H_DE_eff}) has 
the same form as that for the Heisenberg spin model
${\cal H}_{\rm Heis} = -2 \sum_{i<j} J_{ij} \mib{S}_i \cdot \mib{S}_j$
within the leading order of the $1/S$ expansion.
This correspondence gives the relation between
the kinetics of electrons in the DE model and
the exchange coupling $J_{ij}$ in the Heisenberg model,
which is given by Eq.~(\ref{eq:J_ij}).
Thus, up to the leading order of the $1/S$ expansion,
the static part of the magnon self-energy~(\ref{eq:Pi_ij_by_B_ij})
is the same as the effective spin-wave Hamiltonian
for the Heisenberg model.
In other words, 
the spin excitation spectrum of the DE model~(\ref{eq:H_DE})
is equivalent to that of the Heisenberg model
with $J_{ij}$ defined by Eq.~(\ref{eq:J_ij}) in the lowest order of 
$1/S$ expansion.




\begin{thebibliography}{99}

\bibitem{Ramirez1997}
A. P. Ramirez, J. Phys. Cond. Matter. {\bf 9}, 8171 (1997).

\bibitem{Coey1999}
J. Coey, M. Viret, and S. von Molnar, Adv. Phys. {\bf 48}, 167 (1999).

\bibitem{Salamon2001}
M. B. Salamon and M. Jaime, Rev. Mod. Phys. {\bf 73}, 583 (2001).

\bibitem{Imada1998}
M. Imada, A. Fujimori, and, Y. Tokura, Rev. Mod. Phys. {\bf 70}, 1039 (1998).

\bibitem{Tokura1999}
Y. Tokura and Y. Tomioka, J. Mag. Mag. Mater. {\bf 200}, 1 (1999).

\bibitem{Tokura2000}
Y. Tokura and N. Nagaosa, Science {\bf 288}, 462 (2000).

\bibitem{Dagotto2001}
E. Dagotto, T. Hotta, and A. Moreo, Phys. Rep. {\bf 344}, 1 (2001).

\bibitem{Wollan1955}
E. O. Wollan and W. C. Koehler, Phys. Rev. {\bf 100}, 545 (1955).

\bibitem{Searle1970}
C. W. Searle and S. T. Wang, Can. J. Phys. {\bf 48}, 2023 (1970).

\bibitem{Zener1951}
C. Zener, Phys. Rev. {\bf 82}, 403 (1951).

\bibitem{Anderson1955}
P. W. Anderson and H. Hasegawa, Phys. Rev. {\bf 100}, 675 (1955).

\bibitem{deGennes1960}
P. G. de Gennes, Phys. Rev. {\bf 118}, 141 (1960).

\bibitem{Furukawa1999a}
N. Furukawa, {\it Physics of Manganites}, ed. T. A. Kaplan and S. D. Mahanti
(Plenum Press, New York, 1999).

\bibitem{Motome2000}
Y. Motome and N. Furukawa, J. Phys. Soc. Jpn. {\bf 69}, 3785 (2000);
{\it ibid.}, {\bf 70}, 3186 (2001).

\bibitem{MotomePREPRINT}
Y. Motome and N. Furukawa, preprint (cond-mat/0305029).

\bibitem{Hwang1995}
H. Y. Hwang, S.-W. Cheong, P. G. Radaelli, M. Marezio, and B. Batlogg, 
Phys. Rev. Lett. {\bf 75}, 914 (1995).

\bibitem{Radaelli1997}
P. G. Radaelli, G. Iannone, M. Marezio, H. Y. Hwang, S.-W. Cheong, 
J. D. Jorgensen, and D. N. Argyriou, Phys. Rev. B {\bf 56}, 8265 (1997).

\bibitem{Rodriguez-Martinez1996}
L. M. Rodriguez-Martinez and J. P. Attfield, Phys. Rev. B {\bf 54}, 15622 (1996).

\bibitem{Coey1995}
J. M. D. Coey, M. Viret, L. Ranno, and K. Ounadjela, Phys. Rev. Lett. {\bf 75}, 3910 (1995).

\bibitem{Saitoh1999}
E. Saitoh, Y. Okimoto, Y. Tomioka, T. Katsufuji, and Y. Tokura, 
Phys. Rev. B {\bf 60}, 10362 (1999).

\bibitem{Millange1998}
F. Millange, V. Caignaert, B. Domeng\`es, B. Raveau, and E. Suard, 
Chem. Meter. {\bf 10}, 1974 (1998).

\bibitem{Nakajima2002}
T. Nakajima, H. Kageyama, and Y. Ueda, J. Phys. Soc. Jpn.
{\bf 71}, 2843 (2002).

\bibitem{Akahoshi2003}
D. Akahoshi, M. Uchida, Y. Tomioka, T. Arima, Y. Matsui, and Y. Tokura, 
Phys. Rev. Lett. {\bf 90}, 177203 (2003).

\bibitem{Perring1996}
T. G. Perring, G. Aeppli, S. M. Hayden, S. A. Carter, J. P. Remaika, 
and S. -W. Cheong, Phys. Re. Lett. {\bf 77}, 711 (1996).

\bibitem{Moudden1998}
A. H. Moudden, L. Vasiliu-Doloc, L. Pinsard, and A. Revcolevschi, 
Physica B {\bf 241-243}, 276 (1998).

\bibitem{Furukawa1996}
N. Furukawa, J. Phys. Soc. Jpn. {\bf 65}, 1174 (1996).
\label{Furukawa1996}

\bibitem{Hwang1998}
H. Y. Hwang, P. Dai, S.-W. Cheong, G. Aeppli, D. A. Tennant, and H. A. Mook, 
Phys. Rev. Lett. {\bf 80}, 1316 (1998).

\bibitem{Vasiliu-Doloc1998}
L. Vasiliu-Doloc, J. W. Lynn, A. H. Moudden, A. M. de Leon-Guevara, 
and A. Revcolevschi, Phys. Rev. B {\bf 58}, 14913 (1998).

\bibitem{Dai2000}
P. Dai, H. Y. Hwang, J. Zhang, J. A. Fernandez-Baca, S.-W. Cheong, 
C. Kloc, Y. Tomioka, and Y. Tokura, Phys. Rev. B {\bf 61}, 9553 (2000).

\bibitem{Biotteau2001}
G. Biotteau, M. Hennion, F. Moussa, J. Rodr\'{\i}guez-Carvajal, L. Pinsard, 
A. Revcolevschi, Y. M. Mukovskii, and D. Shulyatev, 
Phys. Rev. B {\bf 64}, 104421 (2001).

\bibitem{Furukawa1999b}
N. Furukawa, J. Phys. Soc. Jpn. {\bf 68}, 2522 (1999).

\bibitem{Khaliullin2000}
G. Khaliullin and R. Kilian, Phys. Rev. B {\bf 61}, 3494 (2000).

\bibitem{Solovyev1999}
I. V. Solovyev and K. Terakura, Phys. Rev. Lett. {\bf 82}, 2959 (1999).

\bibitem{Kaplan1997}
T. A. Kaplan and S. D. Mahanti, J. Phys.: Condens. Matter {\bf 9}, L291 (1997).

\bibitem{Golosov2000}
D. I. Golosov, Phys. Rev. Lett. {\bf 84}, 3974 (2000).

\bibitem{Shannon2002}
N. Shannon and A. V. Chubukov, Phys. Rev. B {\bf 65}, 104418 (2002).

\bibitem{Motome2002a}
Y. Motome and N. Furukawa, J. Phys. Soc. Jpn. {\bf 71}, 1419 (2002).

\bibitem{FurukawaPREPRINT}
N. Furukawa and Y. Motome, to appear in Physica B.

\bibitem{Motome2003}
Y. Motome and N. Furukawa, J. Phys. Soc. Jpn. {\bf 72}, 472 (2003).

\bibitem{t_ij}
We consider only positive $t_{ij}$ here.
If some $t_{ij}$ become negative, 
the system suffers from the frustration.
The ground state may be no longer the simple ferromagnetic state
(possibly a glassy state).

\bibitem{Pickett1997}
W. E. Pickett and D. J. Singh, Phys. Rev. B {\bf 55}, 8642 (1997).

\bibitem{Friedel1953}
J. Friedel, Adv. Phys. {\bf 3}, 446 (1953).

\bibitem{Park1998}
J.-H. Park, E. Vescovo, H.-J. Kim, C. Kwon, R. Ramesh, and T. Venkatesan, 
Nature {\bf 392}, 794 (1998).

\bibitem{Furukawa2000}
N. Furukawa and K. Hirota, Physica B {\bf 291}, 324 (2000).

\bibitem{Perring2001}
T. G. Perring, D. T. Adroja, G. Chaboussant, G. Aeppli, T. Kimura, and Y. Tokura, 
Phys. Rev. Lett. {\bf 87}, 217201 (2001)

\bibitem{Kawano1996a}
H. Kawano, R. Kajimoto, M. Kubota, and H. Yoshizawa, 
Phys. Rev. B {\bf 53}, 2202 (1996).

\bibitem{Kawano1996b}
H. Kawano, R. Kajimoto, M. Kubota, and H. Yoshizawa, 
Phys. Rev. B {\bf 53}, 14709 (1996).

\bibitem{Li1997}
Q. Li, J. Zang, A. R. Bishop, and C. M. Soukoulis, 
Phys. Rev. B {\bf 56}, 4541 (1997).

\bibitem{Sheng1997}
L. Sheng, D. Y. Xing, D. N. Sheng, and C. S. Ting, 
Phys. Rev. B {\bf 56}, R7053 (1997).

\bibitem{Goodenough1955}
J. B. Goodenough, Phys. Rev. {\bf 100}, 564 (1955).

\bibitem{DeTeresa1997}
J. M. De Teresa, C. Ritter, M. R. Ibarra, P. A. Algarabel, 
J. L. Garc\'{\i}a-Mu\~noz, J. Blasco, J. Garc\'{\i}a, and C. Marquina, 
Phys. Rev. B {\bf 56}, 3317 (1997).

\bibitem{Tomioka2002}
Y. Tomioka and Y. Tokura, Phys. Rev. B {\bf 66}, 104416 (2002).

\end{thebibliography}
\end{document}